\documentstyle[psfig,a4wide,aps]{revtex} 	% gallery format
\def \Vo {Vorono\"\i\ }
\newtheorem{generic}  {Theorem}
\preprint{HLRZ 65--95}
\psfigurepath{figs}
\begin{document}
\title{Geometrical consequences of foam equilibrium}
\author{C.~Moukarzel~\footnote{Permanent Address: IF-UFF, CEP 24210-340,
Niteroi RJ, Brazil}
}
\address{H\"ochstleistungsrechenzentrum, Forschungszentrum J\"ulich,\\
D-52425 J\"ulich, Germany.}
\date{\today}
\maketitle
\begin{abstract}
The equilibrium conditions impose nontrivial geometrical constraints on
the configurations that a two-dimensional foam can attain. In the first
place, the three centers of the films that converge to a vertex have to
be on a line, i.e. all vertices are \emph{aligned}. Moreover an
equilibrated foam must admit a \emph{reciprocal figure}. This means that
it must be possible to find a set of points $P_i$ on the plane, one per
bubble, such that the segments $\overline{P_i P_j}$ are normal to the
corresponding foam films. It is furthermore shown that these constraints
are  equivalent to the requirement that the foam be a Sectional
Multiplicative \Vo Partition (SMVP). A SMVP is a cut with a
two-dimensional plane, of a three-dimensional Multiplicative \Vo
Partition.  Thus given an arbitrary equilibrated foam, we can always
find point-like sources (one per bubble) in three dimensions that
reproduce this foam as a generalized \Vo partition. These sources are
the only degrees of freedom that we need in oder to fully describe the
foam.
\\
\end{abstract}
%\pacs{PACS numbers:} 
\section{ INTRODUCTION }
\label{sec:introduction}
Cellular structures\cite{Rev1,Rev2,Rev3,Rev4} appear in a wide range of
natural phenomena, and have puzzled and fascinated scientists for
decades\cite{Smith1}. They can be generally described as packings of
space-filling cells of roughly polygonal shape, separated by thin
interfaces to which a surface energy is associated.  They arise as a
consequence of competition between domains under the constraint of
space-filling, and there is often  some mechanism, such as the migration
of a conserved quantity across the interface, which makes them evolve in
time, i.e. coarsen.
\\
Foams such as those obtained by shaking soapy water are  the examples of
cellular structures closest to our everyday experience.  Two-dimensional
foams may be obtained by confining a soap froth between two closely
spaced glass plates. In spite of their apparent simplicity, foams
display much of the phenomenology appearing in coarsening cellular
structures. Foams have been the subject of interest since their
relevance in the problem of grain growth was pointed out by
Smith\cite{Smith2}.  Despite the attention they have received, the
understanding of their dynamical properties proved to be a tricky
problem.  Even some of the most basic issues, such as their asymptotic
scaling properties, has been a matter of debate until recently (see
references above). Numerically exact simulation
procedures\cite{Numerical_Foam}, as well as analytical\cite{Flyvbjerg}
and numerical\cite{Vertex} approximations, have been useful in
understanding the dynamics of ideal foams, but the system sizes
accessible to available simulation procedures are strongly limited (for
a complete account see refs.\cite{Rev3,Rev4}).
\\
A satisfactory theoretical framework for the description of foams has
not yet been achieved. The main result of this work consists in
establishing a \emph{rigorous connection} between foams and \Vo
partitions(VP). This connection provides a set of fundamental degrees of
freedom for the foam (the source's locations in space) and therefore
constitutes a step towards the above mentioned theoretical
understanding. On the other hand, an efficient method for the numerical
simulation of ideal foams would also be highly desirable, and we propose
that such a method could be obtained by exploiting the equivalence
between foams and VP reported in this work.  More precisely, we
establish the following \emph{correspondence} between equilibrated
two-dimensional foams (EF) and a generalization of VPs, the Sectional
Multiplicative \Vo Partition (SMVP):
\\ 
\emph{
Given an arbitrary EF, it is always possible to find sources in space
such that a SMVP with respect to them exactly gives this foam.
}
\\
In Section~\ref{sec:tessellations} the VP and its generalizations are
reviewed.  The simplest generalizations of the VP concept are the
Sectional \Vo Partition and the Multiplicative \Vo Partition. These
correspond to adding a constant to the (square) distance and to
multiplying it by a constant respectively. At the end of this section a
combination of both is introduced, the Sectional Multiplicative \Vo
Partition. We will use this partition in order to describe foams.
\\
The demonstration of the above mentioned correspondence  between foams
and SMVPs is divided in two parts (Sections \ref{sec:SMVPrecognition}
and \ref{sec:equilibrium}) for clarity.  In
Section~\ref{sec:SMVPrecognition}, the \emph{recognition problem} for
SMVPs is solved.  The recognition problem consists in giving the
sufficient geometrical conditions that a circular partition has to
satisfy in order to be a SMVP. We will see that if a circular partition
has all of its vertices \emph{aligned} and admits an \emph{oriented reciprocal
figure}, then it is a SMVP. In other words, it is always possible to
find sources in three-dimensional space that give this circular
partition as a SMVP. In this section we also describe the procedure to
\emph{find} the sources when these conditions are met. Some material
that is needed in this section is described in the appendix.
\\
Section~\ref{sec:equilibrium} deals with the equilibrium conditions for
foams and its geometrical consequences.  We start by writing the force
and pressure equilibrium conditions in compact form in
Section~\ref{sec:eq_equations}.  In Sections~\ref{sec:eq_align} and
\ref{sec:eq_reciprocal} it is shown that an equilibrated foam has all of
its vertices \emph{aligned} and admits an \emph{oriented reciprocal
figure} respectively. Therefore all equilibrated foams satisfy the
conditions required in Section~\ref{sec:SMVPrecognition}, and can be
described as SMVPs. 
\\
Finally in Section~\ref{sec:conclusions} the
implications of this result are discussed, and some perspectives for
future work are advanced.
\section{ Space Partitions}
\label{sec:tessellations}
\subsection{\Vo  Partition }
\label{sec:voronoi}
\begin{figure}[htpb] \vbox{ 
\centerline{\psfig{figure=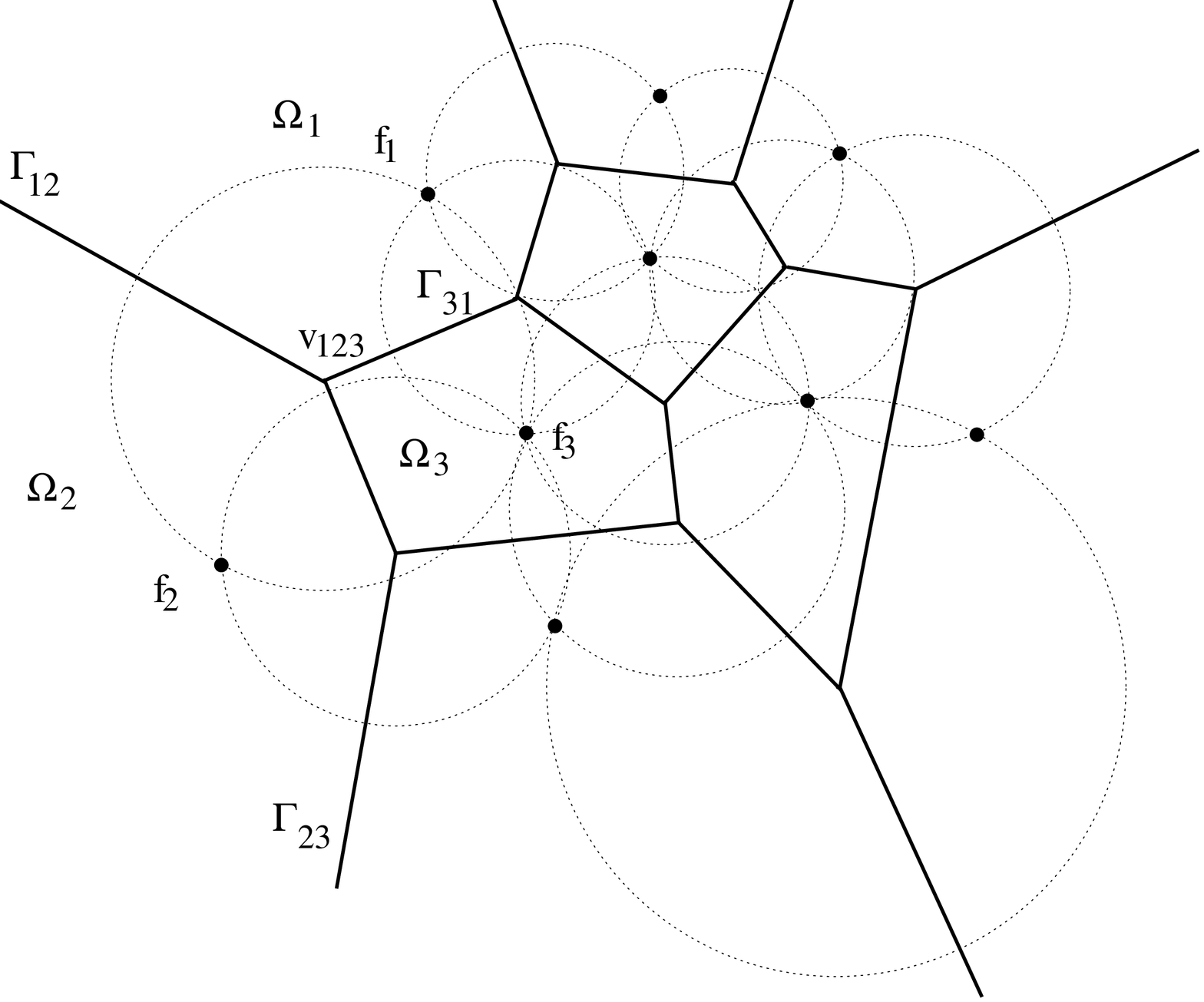,width=11cm,angle=270} }
\centerline{}
\caption{
Shown is a \Vo partition of two-dimensional space with respect to
point-like sources $\{f_i\}$ (black dots). Each source $f_i$ has an
associated cell $\Omega_i$, which is the subset of space that is closer
to $f_i$ than to any other source. Interfaces $\Gamma_{ij}$ (thick
lines) are equidistant from the sources whose cells they delimit.  Three
interfaces $\Gamma_{ij}$, $\Gamma_{jk}$ and $\Gamma_{ki}$ meet at a
vertex $v_{ijk}$, which is therefore the center of a circle (dashed)
trough the three corresponding sources.
}
\label{fig:voronoi_partition}
} \end{figure}

Given a set of $N$ point-like {\it sources}  $\{f_i\}$ in $n$-space,
the {\it \Vo Partition} (VP) or {\it tessellation} of space
with respect to $\{f_i\}$ is a classification of space into {\it cells}
$\Omega_i$ defined such that $x \in \Omega_i$ if $x$ is closer to $f_i$
than to any other source~\cite{Boots_book} .
\begin{equation}
\Omega_i = \{ x \in R^n  / \qquad d(x,i) < d(x,j) \quad \forall j \ne i\}
\label{eq:vorcell}
\end{equation}
where $d(x,i)$ denotes the distance between $x$ and source $f_i$. This
construction is also known under the names:  Wigner-Seitz cells,
Dirichlet Tessellations, Thiessen Polygons. In two dimensions, the dual
lattice of a \Vo partition is called Delaunay triangulation. Cell
$\Omega_i$ can be seen as the ``region of influence'' of $f_i$, were the
sources competing for some spatially distributed resource. It is common
to use these constructions, and their generalizations, as
\emph{approximate} models for cellular structures occurring in
nature\cite{Boots_book,Honda,Icke,Tesis}.
\\
Two neighboring cells $\Omega_i$ and $\Omega_j$ are delimited by an
{\it interface} $\Gamma_{ij}$ of points $x$ equidistant from $f_i$ and
$f_j$.
\begin{equation}
\Gamma_{ij} = \{ x \in R^n  / \qquad d(x,i) = d(x,j) < d(x,k) \quad \forall k \ne i,j\}
\label{eq:vorinterface}
\end{equation}
These interfaces are $(n-1)$-dimensional hyperplanes normal to
$\overline{f_i f_j}$. In two dimensions $\Omega_i$ are convex polygons
and the interfaces $\Gamma_{ij}$ are straight lines
(Fig.~\ref{fig:voronoi_partition}). Three interfaces $\Gamma_{ij}$,
$\Gamma_{jk}$ and $\Gamma_{ki}$  meet at a {\it vertex} $v_{ijk}$, which
is equidistant from $f_i$, $f_j$ and $f_k$. This means that $v_{ijk}$ is
the center of a circle through $f_i$, $f_j$ and $f_k$. Vertices of
higher multiplicity are possible for particular locations of the
sources. For example a fourfold vertex would exist if four sources lay
on the same circle. We will ignore the existence of these particular
configurations, that is we will assume \emph{generic} source locations.
Under this assumption, all vertices are of multiplicity three.
\\
The concept defining \Vo partitions is \emph{equidistance}.  A simple
way to generalize them is changing the way in which distances to the
sources are measured.  One lets each source measure the distance to
points $x$ according to its own ``rule''. The interface $\Gamma_{ij}$ is
then the set of points $x$ for which two neighbors ``claim'' the same
distance.  Two simple ways to do this are: adding an arbitrary constant
to the square of the distance (Sectional \Vo
partition~\cite{Boots_book,Imai,AshBolker2}), and, multiplying the
distance by a constant (Multiplicative \Vo
partition~\cite{Boots_book,Boots,AshBolker2,CFM}).
\subsection{Sectional \Vo Partitions}
\label{sec:sectional}
\begin{figure}[htpb] \vbox{ 
\centerline{\psfig{figure=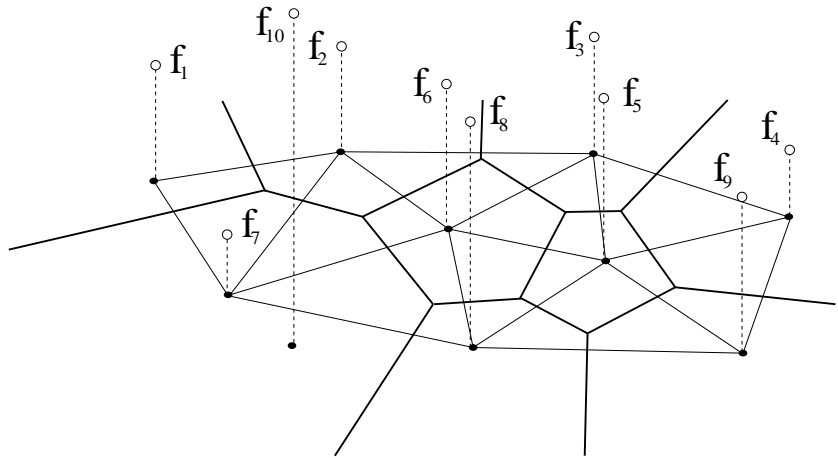,width=11cm,angle=270} }
\centerline{}
\caption{ 
A Sectional \Vo Partition (SVP) of the plane is defined by intersecting
a three-dimensional \Vo Partition with that plane. Sources $P_i$ of this
SVP (black dots) are the projections of the three-dimensional sources
$f_i$ (white dots) onto the plane.   Interfaces $\Gamma_{ij}$ between
cells are rectilinear, normal to $\overline{P_i P_j}$ and meet at triple
vertices $v_{ijk}$, but are \emph{not} equidistant from the sources.
Sources with a too large value of $z_i$  (for example $P_10$ in this
figure) do not have an associated cell.
}
\label{fig:SVPview}
} \end{figure}

A {\it Sectional \Vo Partition} (SVP) is defined~\cite{Boots_book,Imai,AshBolker2}
as a $k$-dimensional cut of a higher dimensional \Vo partition. The
sources $\{P_i\}$ of the sectional partition are defined as the
projections of the original sources $\{f_i\}$ onto this lower
dimensional $k$-space. Cell $\Omega_i$ associated to source $P_i$ is the
intersection of cell $\tilde \Omega_i$ (associated to $f_i$) with this
$k$-dimensional hyperplane. For example, take sources $\{f_i\}$ in three
dimensions (Fig.~\ref{fig:SVPview}) and construct a VP with  three-dimensional cells $\tilde
\Omega_i$ and plane interfaces $\tilde \Gamma_{ij}$.  Now project the
sources  $f_i = (x_i,y_i,z_i)$ onto the $z=0$ plane (which we call the
$\Pi_z$ plane) to obtain the projected sources $P_i = (x_i,y_i)$, which
have associated ``heights'' $z_i$.  Assign to each source $P_i$ the
intersection $\Omega_i$ of $\tilde \Omega_i$ with $\Pi_z$. This defines
the a sectional partition of  $\Pi_z$ with respect to sources $\{P_i\}$ with
heights $z_i$. Cells $\Omega_i$ are thus defined as
\begin{equation}
\Omega_i = \{ x \in \Pi_z  / \qquad d^2(x,i) + z_i^2 < d^2(x,j) + z_j^2
\quad \forall j \ne i\}
\label{eq:svpcell}
\end{equation}
\noindent 
where $d(x,i)$ is the distance, on the $\Pi_z$ plane, between
$P_i$ and point $x$. 
\\
In the same way, interfaces $\Gamma_{ij}$ are defined as
\begin{equation}
\Gamma_{ij} = \{ x \in \Pi_z  / \qquad d^2(x,i) + z_i^2 = d^2(x,j) +
z_j^2 < d^2(x,k) + z_k^2 \quad \forall k \ne i,j\}
\label{eq:svpinterface}
\end{equation}
Properties of these partitions in two dimensions are:
\begin{itemize}
\item Interfaces are straight lines and (generically) meet at triple
vertices. ( In dimension $d$, vertices have multiplicity $d+1$).
\item Interface $\Gamma_{ij}$ is normal to $\overline{P_i P_j}$, but
not in general equidistant from $P_i$ and $P_j$. 
\item The partition is unchanged if all $z_i$ are changed according to:
$z_i ^2 \to z_i^2 + c^2$ for $c$ an arbitrary real.
\end{itemize}
An important difference with the \Vo partition, is that in this case
there may be sources to which no cell is associated. This happens if
$\tilde \Omega_i$ is not cut by $\Pi_z$.
It is easy to see that  SVPs are \emph{equivalent} to Laguerre Partitions~\cite{Imai}.
\\
Appendix~\ref{sec:RPrecognition} discusses the \emph{recognition
problem} for this kind of partitions, that is giving the sufficient
conditions for an arbitrary rectilinear partition to be a SVP. We will
use similar concepts in Section~\ref{sec:SMVPrecognition} in order to
solve the recognition problem for Sectional Multiplicative \Vo
Partitions.
\\
As an example of the application of sectional partitions, consider the
case of sources $f_i$ whose location in space changes in time, giving
rise to a time-dependent partition of $\Pi_z$. If for example source
$f_1$ moves away from $\Pi_z$, the two-dimensional cell $\Omega_1$ will
shrink and finally disappear when $\tilde \Omega_1$ no longer cuts
$\Pi_z$. Therefore the number of cells can vary without changing the
number of sources. Thus sectional partitions can be used as dynamical
models for crystal growth~\cite{Tesis}, and other processes in which
some cells disappear or are created.
\subsection{Multiplicative \Vo Partition}
\label{sec:mvp}
\begin{figure}[htpb] \vbox{ 
\centerline{\psfig{figure=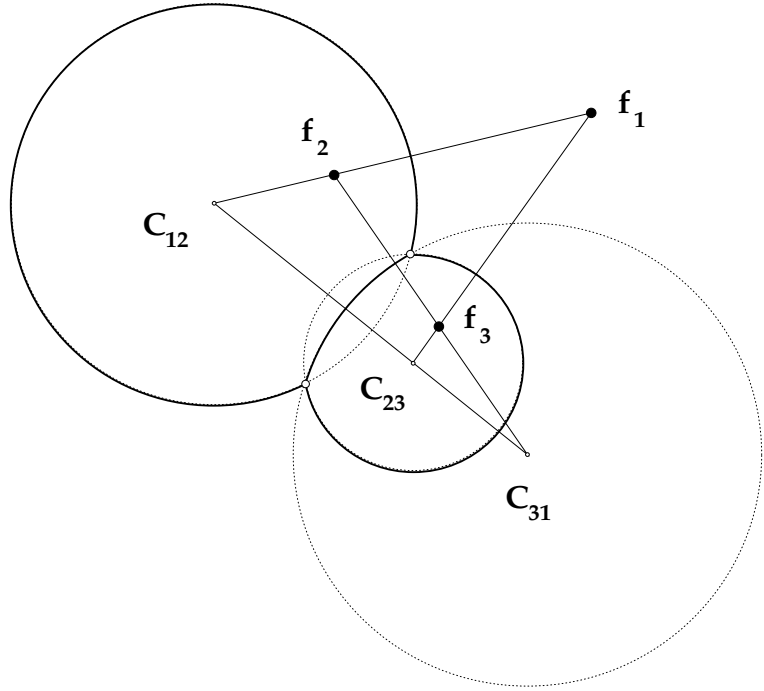,width=10cm,angle=270} }
\centerline{}
\caption{
A Multiplicative \Vo Partition with respect to three sources in two
dimensions. The source with the largest intensity $a_i$ (source $f_1$ in
this figure) has an unbounded cell. The other two sources are within
their associated cells.  Angles formed by the interfaces at the vertices
are depend on the positions of the sources and the values of the
intensities. For certain values of $\{a_i\}$ the interfaces may not
intersect. In this case the partition would be composed of two disjoint
circles containig one of the sources with smaller intensities each. The
MVP with respect to three sources in three-dimensional space is a
cluster of two spherical bubbles (Fig.~\ref{fig:smvp_vertex}, dashed
lines).
}
\label{fig:mvp_vertex_2d}
} \end{figure}

The Multiplicative \Vo Partition (MVP)~\cite{Boots,AshBolker2,CFM} is
defined by assigning to each source $f_i$ a positive  {\it intensity}
$a_i$, and defining the \emph{multiplicative distance}
$d_m(x,i)=d(x,i)/a_i$. The cell $\Omega_i$ associated to source $f_i$ is
then the set of points $x$ closer to $f_i$ (in terms of this
multiplicative distance) than to any other source
\begin{equation}
\Omega_i = \{ x \in R^n  / \qquad {d(x,i)\over a_i}  < {d(x,j)\over a_j}
\quad \forall j \ne i\}
\label{eq:mvpcell}
\end{equation}
Interfaces $\Gamma_{ij}$ are hyperspherical surfaces
(circle arcs in two dimensions) satisfying
\begin{equation}
\Gamma_{ij} = \{ x \in R^n  / \qquad {d(x,i)\over a_i}  = {d(x,j)\over
a_j} < {d(x,k)\over a_k}\quad \forall k \ne i,j\}
\label{eq:mvpinterface}
\end{equation}
In two dimensions, the circular interface of a MVP with two sources is
one of the Apollonius circles (see for example~\cite{Coxeter}) of those
two points. In arbitrary dimensions let $d_{ij}$ be the distance between
two sources $f_i$ and $f_j$, and $A_{ij} = a_j/a_i$.  Without loss of
generality we take $A_{ij}<1$.  This means that $f_j$ has the smallest
intensity and therefore $\Omega_j$ will be the interior of a
hypersphere, while $\Omega_i$ will be its exterior. The following
properties are satisfied in any dimension.
\begin{itemize}
\item $f_i$ is contained in $\Omega_i$.
\item $\Gamma_{ij}$ has radius \hbox{$ R_{ij} =  A_{ij} d_{ij} / ( 1 -
A_{ij}^2 ) $}, and its center $C_{ij}$  is on the straight line going
through $f_i$ and $f_j$.
\item $C_{ij}$ is located at a distance $ d_{ij} / (1-A_{ij}^2)$ from
$f_i$, that is, it never lays between $f_i$ and $f_j$.
\end{itemize}
Three sources give rise to a tessellation like the one shown in
Fig.~\ref{fig:mvp_vertex_2d}. The exterior of the two ``bubbles'' is the
infinite cell associated to the source with the largest intensity, $f_1$
in this example. For some choices of $\{a_i\}$ the interfaces will not
intersect. In this case the tessellation is simply a pair of circles,
each containing one of the sources with smaller intensities, while the
exterior of these two circles is the cell of the source with larger
intensity. MVPs can be interpreted again as ``areas of influence'' of
sources $f_i$, but now each source has a different strength.
\\
Consider a  MVP of  two dimensional space with respect to a set of
sources $\{f_i\}$. If three interfaces $\Gamma_{ij}$, $\Gamma_{jk}$ and
$\Gamma_{ki}$ meet at a vertex (again, vertices of higher multiplicity
are only possible for particular configurations, which we ignore here),
then centers $C_{ij}$, $C_{jk}$ and $C_{ki}$ lay on a line. The reason
why this is so is simple. If two interfaces $\Gamma_{ij}$ and
$\Gamma_{jk}$ have a common point $v_{ijk}$ , then their
\emph{continuations} must meet again at another point $v^*_{ijk}$, the
\emph{conjugate vertex}. But then the third interface $\Gamma_{ik}$ must
also go through this point since $d(v^*,i)/a_i=d(v^*,j)/a_j$ and
$d(v^*,j)/a_j=d(v^*,k)/a_k$ implies $d(v^*,i)/a_i=d(v^*,k)/a_k$.
Therefore two points exist ($v$ and $v^*$) at which all three circles
intersect. Then the centers of these circles must be on a line
(Fig.~\ref{fig:mvp_vertex_2d}).
\\
Let us define a  \emph{Circular Partition}(CP) of two-dimensional space
to be a classification of space in cells delimited by arbitrary circle
arcs that meet at triple points called vertices. No three centers of
these arbitrarily defined interfaces will in general be on a line. We
will say that a vertex of a CP is \emph{aligned} if the centers of the
three interfaces defining it are on a line.
As we saw, all vertices of a MVP are aligned in 2d.  Therefore for each
triple vertex of a MVP in 2d, the three sources and the three centers
form the configuration~\cite{ProjectiveGeometry} $(6_2,4_3)$ of
projective geometry. Since $C_{ij}$ cannot be between $P_i$ and $P_j$,
the three centers can only be on one of the two external segments of the
configuration.
\\
In a three-dimensional MVP, four cells meet at each vertex $v_{ijkl}$,
giving rise to six spherical interfaces $(ij), (ik), (il), (jk), (jl),
(kl)$. The centers of these interfaces are aligned in triplets so that
the six centers also form the configuration $(6_2,4_3)$, this time in
three-dimensional space.  This has the implication that the six centers
are necessarily on the same plane.
\\
The MVP was first introduced by Boots~\cite{Boots} to describe areas of
influence in geography. Ash and Bolker~\cite{AshBolker2} also studied
the recognition problem, that is, under which conditions a CP is a MVP.
The visual similarity between this kind of partitions and
two-dimensional foams is striking, and suggests the idea to find a
connection between them.  Obviously the centers of the films of a foam
must be aligned for each triple vertex, if the foam is to be described
as a MVP, since MVPs are aligned. One finds~\cite{CFM} that this
alignment condition is indeed satisfied by all vertices of arbitrary
equilibrated foams. Despite this (which is a necessary but not
sufficient condition for a foam to be a MVP), two-dimensional MVPs
cannot describe all possible equilibrated foams in two dimensions. This
work shows the reason of this limitation: In order to describe arbitrary
equilibrated two-dimensional foams we must introduce the sectional
variant of a multiplicative partition. In other words, instead of
confining the sources to the plane, we let them exist in a
three-dimensional space and obtain the foam as \emph{a two-dimensional
cut of a three-dimensional MVP}.
\subsection{Sectional Multiplicative \Vo Partition}
\label{sec:smvp}
\begin{figure}[htpb] \vbox{ 
\centerline{\psfig{figure=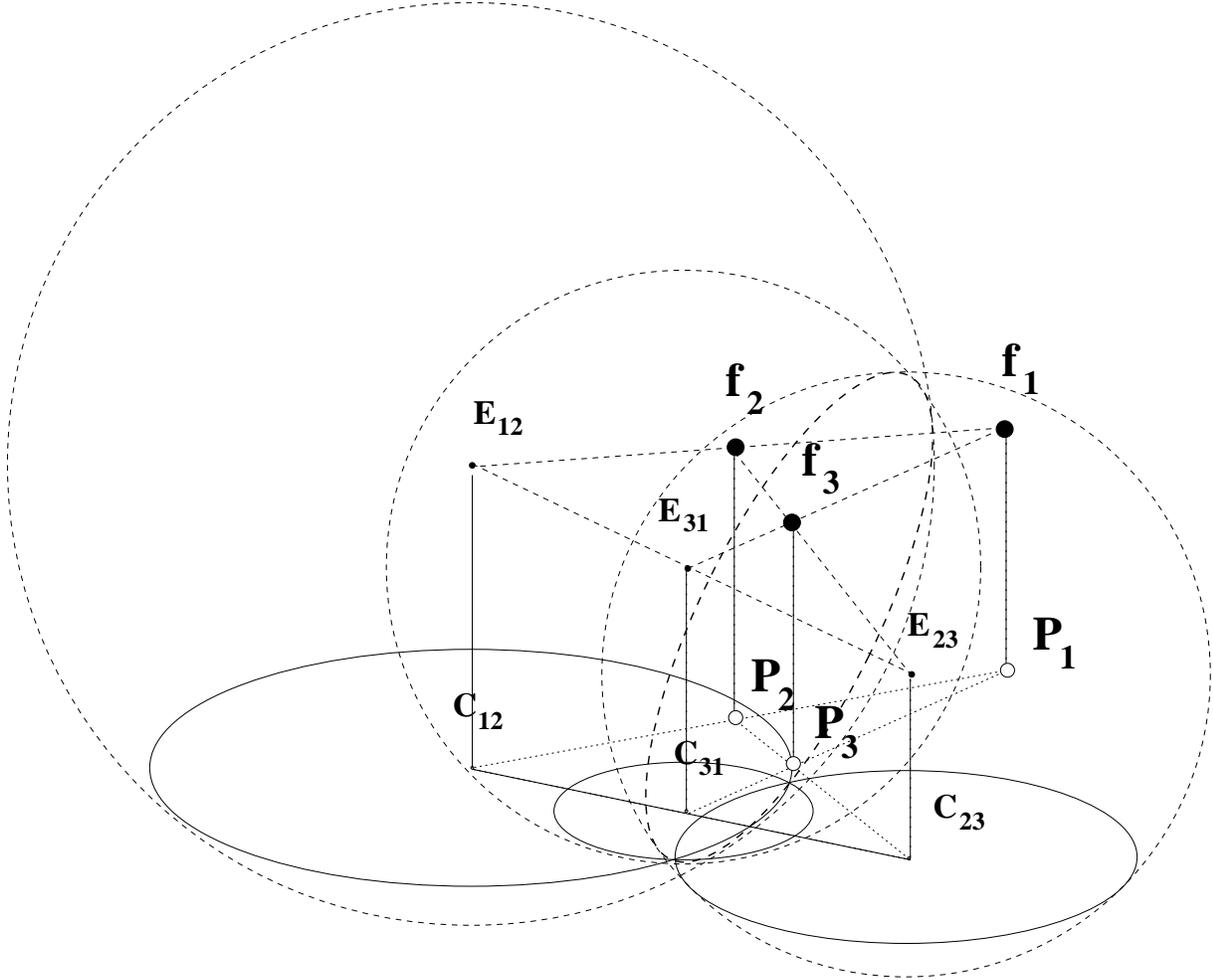,width=16cm,angle=270} }
\centerline{}
\caption{ 
A Sectional Multiplicative \Vo Partition (SMVP) is defined as a plane
section (black circles) of a three-dimensional Multiplicative \Vo
Partition (dashed lines). Here an example for three sources. Sources
$P_i$ of the SMVP (on the $\Pi_z$ plane, white dots), are the
projections of sources $f_i$ of the MVP in three dimensions (black
dots). The centers $E_{ij}$ of the 3d MVP are aligned, which implies the
alignement of the centers $C_{ij}$ of the SMVP. Segments $\overline{f_i
f_j}$ are normal to spheres $\tilde \Gamma_ij$, therefore $\overline{P_i
P_j}$ are normal to circles $\Gamma_{ij}$ on the plane.
 }
\label{fig:smvp_vertex}
} \end{figure}

\begin{figure}[htpb] \vbox{ 
\centerline{\psfig{figure=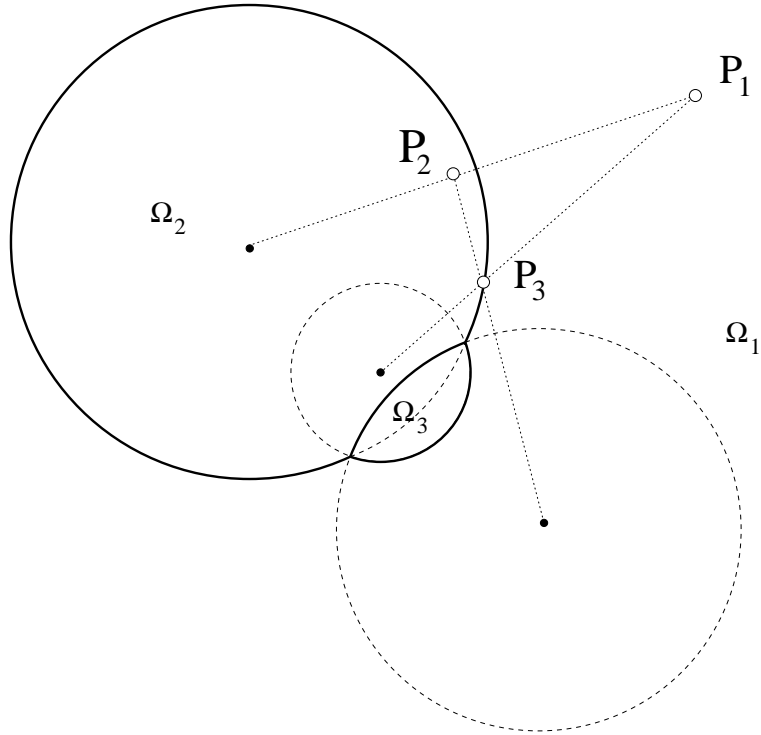,width=10cm,angle=270} }
\centerline{}
\caption{ 
An upper view of the SMVP with respect to three sources defined in
Fig.~\ref{fig:smvp_vertex}.  Notice that sources $P_i$ are not all
contained in their cell $\Omega_i$, as happens in the case of the MVP
(Fig.~\ref{fig:smvp_vertex}).
}
\label{fig:smvp_upper}
} \end{figure}

\begin{figure}[htpb] \vbox{ 
\centerline{\psfig{figure=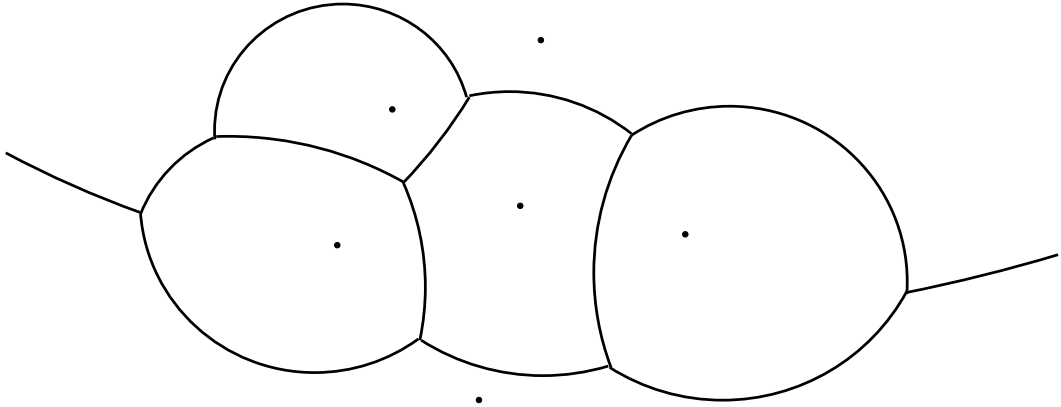,width=14cm,angle=270} }
\centerline{}
\caption{ 
A SMVP of the plane with six sources. The projections $P_i$ of the
sources are indicated as small dots in the figure. For each one of the
vertices in this partition, the three interfaces can be continued and
they will meet each other again at a \emph{conjugate}vertex.  
}
\label{fig:mvp2d-many}
} \end{figure}

We will restrict the description to the case of a plane
cut of a three dimensional MVP. The generalization to a $k$-dimensional
cut of an $n$-dimensional MVP is straightforward.
\\
A plane cut, with a plane $\Pi_x$, of a three-dimensional  MVP defines a
Sectional Multiplicative \Vo Partition (SMVP) of $\Pi_z$ (See
Figs.~\ref{fig:smvp_vertex} and \ref{fig:smvp_upper}). The sources $P_i$
of this SMVP are the normal projections onto $\Pi_z$ of the original
sources $f_i$, and may be seen to have as attributes both an intensity
$a_i$ and a height $z_i$. Cells  $\Omega_{i}$ associated to sources
$P_i$ are defined as
\begin{equation}
\Omega_i = \{ x \in \Pi_z  / \qquad {d(x,i)^2 + z_i^2 \over a_i^2}  <
{d(x,j)^2 + z_j^2 \over a_j^2} \quad \forall j \ne i\}
\label{eq:smvpcell}
\end{equation}
In the same way, interfaces $\Gamma_{ij}$ are circle arcs satisfying 
\begin{equation}
\Gamma_{ij} = \{ x \in \Pi_z  / \qquad 
{ d^2(x,i) + z_i^2 \over a_i^2 }= { d^2(x,j) + z_j^2 \over a_j^2 }
< { d^2(x,k) + z_k^2 \over a_k^2 } \qquad \forall k \ne i,j \}
\label{eq:smvpinterface}
\end{equation}
An example of a SMVP with three sources is shown in
Figs.~\ref{fig:smvp_vertex} and \ref{fig:smvp_upper}. We see that there
are two vertices on $\Pi_z$ at which the three interfaces meet. In a
general case (for example in a partition with respect to many sources
like in Fig.~\ref{fig:mvp2d-many}), if vertex $v_{ijk}$ exists, then the
continuations of interfaces $\Gamma_{ij}$, $\Gamma_{jk}$ and
$\Gamma_{ki}$ meet at a {\it conjugate} vertex $v^*_{ijk}$ also. This
implies the alignment of centers $C_{ij}$, $C_{jk}$ and $C_{ki}$, which
could also be deduced from the alignment of  $E_{ij}$, $E_{jk}$ and
$E_{ki}$ in three-dimensions (Fig.~\ref{fig:smvp_vertex}) .  \\ We
notice that the SMVP is equivalent to a Multiplicative Laguerre
Partition, since a SVP is equivalent to a Laguerre
Partition~\cite{Imai}.  As is usual in sectional partitions, in the SMVP
there may be sources $P_i$ with no associated cell, those for which the
corresponding three-dimensional ``bubble'' $\tilde \Omega_i$ does not
cut $\Pi_z$.
\\ 
It was seen in section~\ref{sec:mvp} that the spherical interface
$\tilde \Gamma_{ij}$ in three dimensions is cut normally by the segment
$\overline{f_i f_j}$.  As a projective consequence of this, the straight
line containing  $P_i$ and $P_j$ on $\Pi_z$ is normal to the circular
interface $\Gamma_{ij}$.  In other words, $P_i$, $P_j$ and $C_{ij}$ are
on the same line.  Furthermore and since $E_{ij}$ never lays between
$f_i$ and $f_j$ in 3d, we notice that $C_{ij}$ is always outside the
segment $\overline{P_i P_j}$.
\section{When a Circular Partition is a SMVP}
\label{sec:SMVPrecognition}
In this section the sufficient conditions for a CP to be a SMVP are
given. We first introduce the notion of oriented reciprocal figure
for CPs, and then proceed to demonstrate that an aligned CP admitting
such a reciprocal figure is a SMVP.
\subsection{Reciprocal figure of a circular partition}
\label{sec:CPreciprocalfig}
We will now generalize the concept of reciprocal figure as appropriate
for circular partitions.  We say that a graph $\cal R$ made of points
$P_i$ connected by edges $(ij)$ forms a reciprocal figure for a CP $\cal
P$ if,
\begin{itemize}
\item points $P_i$ are in correspondence with the cells $\Omega_i$ of
$\cal P$.
\item edges $(ij)$ are in correspondence with the interfaces $\Gamma_{ij}$ of
$\cal P$.
\item for each $\Gamma_{ij}$ in $\cal P$, points $P_i$, $P_j$ and
$C_{ij}$ lay on the same line. 
\end{itemize}
Consider two regions $\Omega_i$ and $\Omega_j$ separated by a circular
interface $\Gamma_{ij}$. A reciprocal figure $\cal R$ will be said to respect
\emph{orientation} if for each $\Gamma_{ij}$ in $\cal P$:
\begin{itemize}
\item $C_{ij}$ is not between $P_i$ and  $P_j$.
\item starting from $C_{ij}$ and traveling along $\overline{C_{ij} P_i
P_j}$ to infinity, points $P_i$ and $P_j$ are found in the same order
as regions $\Omega_i$ and $\Omega_j$ are.
\end{itemize}
We saw already that all SMVPs are aligned (all vertices satisfy
the alignment condition introduced in Sec.~\ref{sec:mvp}). On the other
hand, it is clear that the sources $P_i$ of a SMVP form a reciprocal
figure that satisfies orientation. Therefore
$$
\cal P  \hbox{ is a SMVP } \Rightarrow \cal P \hbox{ is aligned and
admits an oriented reciprocal figure}
$$
Figure~\ref{fig:convexity} shows four possible partitions that share the
same centers $C_{12}$, $C_{23}$ and $C_{31}$.  All four admit reciprocal
figures. But only in cases a) and c) a reciprocal figure that satisfies
orientation can be constructed. Therefore neither b) nor c) can be a
SMVP. The reason is that the orientation condition cannot be satisfied
if any of the vertices is not \emph{convex}. A vertex is \emph{convex}
if all the internal angles formed by the interfaces are less than $\pi$.
Convexity of all vertices is clearly a necessary condition for a
circular partition to be a SMVP.
\\
We are now ready to  give the \emph{sufficient} conditions for a
circular partition to be a SMVP.  We will show in
Section~\ref{sec:find_sources} that if a circular partition $\cal P$ is
aligned and admits a reciprocal figure that satisfies orientation, then it
is a SMVP.
$$
\cal P  \hbox{ is aligned  and admits an oriented reciprocal figure}
\Rightarrow \cal P \hbox{ is a SMVP}
$$
\subsection{Finding the sources of a SMVP.}
\label{sec:find_sources}
\begin{figure}[htpb] \vbox{ 
\centerline{\psfig{figure=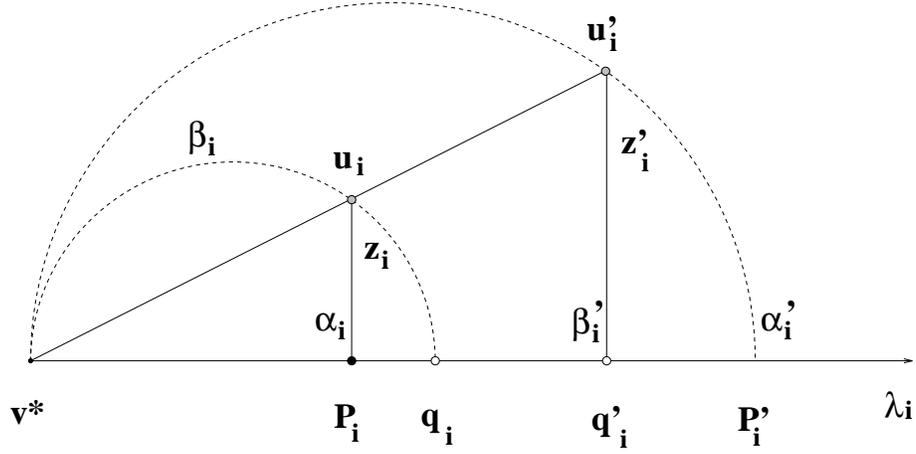,width=12cm,angle=270} }
\centerline{}
\caption{
Shown is a plane  normal to $\Pi_z$ and containing $\lambda_i$.
This figure shows how the height $z_i$ of source $f_i=u_i$ in the original
system is determined, given the positions of $P_i$ and $q_i$.  Primed
variables correspond to positions after the inversion, that is, in the
``straight'' system. Sources $f_i'=u_i'$ are above $q_i'$, and therefore the
back-transformed sources $f_i$ are above $P_i$.
}
\label{fig:above}
} \end{figure}

\begin{figure}[htpb] \vbox{ 
\centerline{\psfig{figure=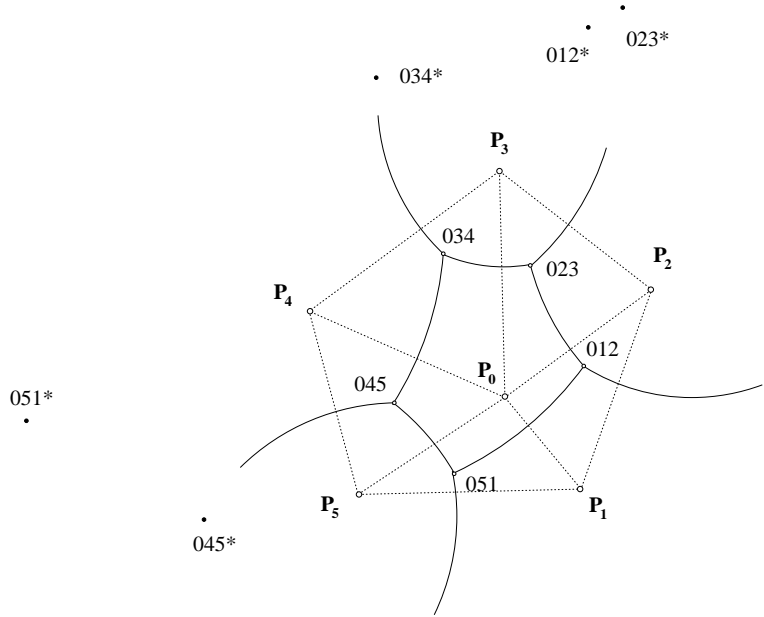,width=10cm,angle=270} }
\centerline{}
\caption{
The points $P_i$ of the reciprocal figure (empty circles) are the
projections, on the $xy$ plane, of the sources $f_i$ in
three-dimensional space. The lines $\overline{P_i P_j}$ must  thus be
normal to  interfaces $\Gamma_{ij}$. Black dots indicate the location of
the conjugate vertices $v^*_{ijk}$ for each vertex $v_{ijk}$ in the
figure.
}
\label{fig:bubblesources}
} \end{figure}

Given a circular partition $\cal P$ of a region of two dimensional
space, all whose (triple) vertices are aligned, we show here that, if
$\cal P$ admits a reciprocal figure that satisfies orientation, then
$\cal P$ is a SMVP with respect to sources $f_i$ located somewhere above
points $P_i$. Our demonstration is constructive, that is we explicitly
show how the sources and intensities are determined. We will use for
this purpose the inversion transformation (Appendix~\ref{sec:inversion})
to \emph{straighten} each vertex in turn, that is, transform a circular
vertex into a rectilinear one. This rectilinear vertex is one of a of a
SVP.  Sources are located in this \emph{straight} representation
(Appendix~\ref{sec:RPrecognition}), and then transformed back to the
original system. The intensities $a_i$ of the corresponding
SMVP are obtained in this back-transformation, since the SMVP is
invariant under inversion.
\\
Let $v$ be an aligned vertex on which three interfaces $\{\Gamma_{12},
\Gamma_{23}, \Gamma_{32}\}$ meet, and $\{ P_1, P_2, P_3 \}$  the three
points  of the reciprocal figure associated to the three bubbles sharing
the vertex, as in Fig.~\ref{fig:reciproca1}. The conjugate vertex $v^*$
is obtained as the intersection point of the \emph{continuations} of the
interfaces, which happens at a unique point because of alignment. Let
furthermore $\lambda_i$ be the straight line through $v^*$ and $P_i$.
We start by showing that
\begin{generic} 
\label{th:recircles}
A monoparametric family of triplets of circles $\{ \omega_{12},
\omega_{23}, \omega_{31} \}$ exists, which has the following
properties:
\begin{itemize}
\item $\omega_{12}$, $\omega_{23}$ and $\omega_{31}$ intercept each
other on  $v^*$.
\item $\omega_{ij} \perp \Gamma_{ij}$.
\item The intersection point $q_i$ between $\omega_{ij}$ and
$\omega_{ik}$ lays on $\lambda_i$.
\end{itemize}
\end{generic}
The non-trivial content of the theorem is the fact that the three
intersection points $q_i$ of these normal circles $\omega_{ij}$ lay on
the lines $\lambda_i$.  For example take an arbitrary point $q_1$ on
$\lambda_1$ and draw a circle $\omega_{12}$ through $q_1$, $v^*$ and
normal to $\Gamma_{12}$.  Let $q_2$ be the intersection of $\omega_{12}$
with $\lambda_2$. Draw now a circle $\omega_{23}$ trough $q_2$, $v^*$
and normal to $\Gamma_{23}$. Let $q_3$ be its intersection with $\lambda_3$.
Then the circle $\omega_{31}$ through $v^*$, $q_1$ and $q_3$ is normal
to $\Gamma_{13}$. 
\\
\noindent {\bf  Proof.} The demonstration is done by first performing an
inversion with center $v^*$ and arbitrary radius, whereupon the
interfaces $\Gamma_{ij}$ are transformed into straight lines
$\Gamma_{ij}'$ meeting at $v'$ (Fig.~\ref{fig:reciproca2}). This
transformed system will be referred to as the ``straight''
representation of that vertex.  Primed names are used in the straight
representation, with the exception of $v^*$, whose original location is
kept in Fig.~\ref{fig:reciproca2} ($v^*$ itself is mapped to $\infty$ by
the inversion). Lines originally not going through $v^*$ are transformed
into circles through $v^*$, as is the case of $\{S_{12}, S_{23},
S_{31}\}$, which form the reciprocal figure in the original system.
Circles $\omega_{ij}$ are transformed into straight lines $\omega_{ij}'$
going through $q_i'$ and $q_j'$, respectively on $\lambda_i$ and
$\lambda_j$, which are invariant.
\\
The inversion transformation preserves angles, therefore circle
$S_{ij}'$ is normal to the now straight interface $\Gamma_{ij}'$. This
means that its center $L_{ij}$ must lay on $\Gamma_{ij}'$. Consider the
figure formed by $v'$, $L_{12}$, $L_{23}$ and $L_{31}$, and the six
lines joining them. Maxwell~\cite{Maxwell1} showed that a figure made
of four points  joined by  six lines \emph{always} has a reciprocal
figure (to see this just consider the centers of the four circles going
through three of these points). We will now identify it on
Fig.~\ref{fig:reciproca2} as the figure formed by $v^*$, $q'_1$, $q'_2$
and $q'_3$.  First we notice that circles $S_{ij}'$ and $S_{jk}'$
intercept each other at two points $v^*$ and $P_j'$, both on
$\lambda_j$. Therefore lines $\overline{L_{ij} L_{jk}}$ are normal to
$\lambda_j$, and we have identified the first three lines of the
reciprocal figure.  The other three lines $\omega_{ij} = \overline{q_i'
q_j'}$ of the reciprocal figure have to be normal to the segments
$\overline{v L_{ij}}$, so they are normal to the interfaces
$\Gamma_{ij}'$. Since the inversion preserves angles, this means that
$\omega_{ij} \perp \Gamma_{ij}$ in the original system
(Fig.~\ref{fig:reciproca1}).    \hfill $\diamond$ 
\\ \vskip 0.2cm
We see that the existence of triplets of circles $\{\omega_{12},
\omega_{23}, \omega_{31} \}$ with the above mentioned properties is a
consequence of the existence, in the straight representation, of a
figure $v^* q_1' q_2' q_3'$, which is the reciprocal of {$v L_{12}
L_{23} L_{31}$}. In other words, Theorem~\ref{th:recircles} means that
in the straight representation, triangle $q_1' q_2' q_3'$ forms a
reciprocal figure for the rectilinear vertex. Clearly this reciprocal
figure \hbox{$q_1' q_2' q_3'$}  satisfies orientation if the original
figure \hbox{$P_1  P_2  P_3 $} satisfies orientation in the circular
system.  Therefore as discussed in Appendix~\ref{sec:RPrecognition}, it
is always possible to find three sources $f_i', \quad i=1,2,3$, located
at heights $z_i'$ above points $q_i'$, such that this rectilinear vertex
is a SVP with respect to them. Once these sources $f_i'$ are known in
the straight representation, a second inversion around the same point
$v^*$ takes us back to the original system and provides the original
locations $f_i$ of the sources.
\\
If $v_{123}$ is the first vertex to be considered in the system, we have
one degree of freedom in the determination of the circles $\omega_{ij}$,
or equivalently of points $q_i$. Once in the straight representation,
there is one more degree of freedom: the determination of \emph{one} of
the heights $z_i'$. We  will now show that, if this $z_i'$ is chosen
appropriately, the three back-transformed sources $f_i$ are located
above their respective points $P_i$.
\\
Consider a plane $\Pi_i$ that is normal to $\Pi_z$ and contains
$\lambda_i$, as shown in Fig.~\ref{fig:above}. Draw a circle $\beta_i$
through $q_i$, $v^*$ and normal to $\Pi_z$. Let $u_i$ be the
intersection of this circle with the normal $\alpha_i$ to $\Pi_z$
through $P_i$. Now under an inversion with center $v^*$, point $u_i'$
is above $q_i'$, since circle $\beta_i$ is transformed in a straight
line $\beta_i'$, normal to $\Pi_z$. Straight line $\alpha_i$ normal to
$\Pi_z$ through $P_i$ is now a circle $\alpha_i'$ through $v^*$, $u_i'$
and normal to $\Pi_z$. Its intersection with $\Pi_z$ determines the
image $P_i'$.  Now we now that source $f_i'$ is located somewhere on
$\beta_i'$. The sources can be displaced vertically (simultaneously)
according according to what we saw in Appendix~\ref{sec:RPrecognition},
but not independently. Fixing the height $z_i'$ of one of them
determines the other two uniquely. What we want to demonstrate is
that if one of the sources $f_i'$ coincides with its point $u_i'$, then
the other two also do. In order to do this, take for example the sphere
$\tilde {\cal X}_{12}$ through $v^*$, $f_1'$, $f_2'$ and normal to
$\Pi_z$. In the straight representation, interface $\tilde
\Gamma_{12}'$ of the \Vo Partition with respect to sources
$f_1'$ and $f_2'$ is a plane, equidistant from $f_1'$ and $f_2'$, and
normal to $\overline{f_1' f_2'}$.  This means that $\tilde
\Gamma_{12}'$ is normal to $\tilde {\cal X}_{12}$. As a consequence,
the intersection ${\cal X}_{12}$(a circle) of $\tilde {\cal X}_{12}$
with $\Pi_z$  is normal to the intersection $\Gamma_{12}'$(a straight
line) of  $\tilde \Gamma_{12}'$ with $\Pi_z$. Now assume that $f_1' =
u_1'$. This means that $P_1' \in \tilde {\cal X}_{12}$ since
$\alpha_1'$ is in this case contained in $\tilde {\cal X}_{12}$. But
then the circle ${\cal X}_{12}$ is the circle  $S_{12}'$
(Fig.~\ref{fig:reciproca2}) that goes through $P_1'$, $P_2'$ and $v^*$
and is normal to $\Gamma_{12}'$. This implies that $P_2'$ is also in
$\tilde {\cal X}_{12}$, which in turn implies $f_2' = u_2'$. The same
reasoning can be repeated for the pair of sources $1$ and $3$. 
\\
We have thus shown that all sources $f_i$ are, in the back-transformed,
original system, located above the points $P_i$ of the reciprocal
figure. In other words, that the given reciprocal figure is the
projection of the sources $f_i$ on the $\Pi_z$ plane.
\\
Now we have to determine the heights $z_i$ and intensities
$a_i$. The heights $z_i$ of the sources can be found found by
back-transforming the sources $f_i'$ from the straight system. But in
practice this is not necessary. By looking at Fig.~\ref{fig:above}
we notice that  triangle \hbox{$v^* f_i q_i$} is rectangle. Therefore
\begin{equation}
z_i^2	=  | v^* P_i | \times   | P_i q_i |,
\label{eq:zi}
\end{equation}
which suffices to locate the sources $f_i$ with the sole knowledge of
points $P_i$ and $q_i$ in the original system.  Knowing now the spatial
locations of the sources $f_i$ in three dimensions, the intensities
$a_i$ follow immediately since the vertex $v^*$ is
equidistant~\cite{equidistance} from the three sources.
This means
\begin{equation}
a_i  =  A |v^* f_i |  = A ( | v^* P_i| \times | v^* q_i| )^{1/2} 
\label{eq:ai}
\end{equation}
Here $A$ is an arbitrary positive constant.  Alternatively we can get
the same result by using the transformation properties
(eq.~\ref{eq:dist_transf}) of the $a_i$'s and the fact that the
intensities are all equal in the straight system.
\\
We have used the inversion transformation to demonstrate that the given
vertex is a SMVP with respect to sources $f_i$ located above $P_i$, and
showed how the sources $f_i$ can be located without the need to
explicitly perform an inversion for each vertex. All we need is to
know the location of points $q_i$, which we are free to choose for the
first vertex under consideration. This first choice determines all
subsequent $q_i$ points, and therefore all sources. It is clear from
Fig.~\ref{fig:above} that $|v^* q_i| \ge |v^* P_i|$ has to be
satisfied. Therefore one has to choose the triplet of circles $w_{ij}$
such that this condition is verified for the three points $q_i$. If one
of the $q_i$ coincides with $P_i$, this means that the corresponding
height $z_i$ is zero, i.e. the source $f_i$ lays on the $\Pi_z$ plane.
A point $q_i$ closer to $v^*$ than the corresponding $P_i$ is not
acceptable.
\\
The general procedure to determine the sources can now be described.
Assume we are given a circular tessellation which satisfies the
required conditions of alignment and convexity, and that we are able
to find, or are given, a reciprocal figure $P_i$ for it. We start from
one of the cells of the partition, as in Fig.~\ref{fig:bubblesources}.
Take an arbitrary vertex to start with, for example vertex $v_{012}$, and
determine tentative locations for point $q_0$ on $\lambda_0$. This
fixes $q_1$ and $q_2$ as discussed in Section~\ref{sec:find_sources}.
After fixing $q_0$ the rest of the construction is uniquely
determined.  The locations  $q_0, q_1, q_2$ determine $z_0, z_1$ and
$z_2$, and the corresponding intensities $a_0, a_1$ and $a_2$, through
equations (\ref{eq:zi}) and (\ref{eq:ai}). Now proceed to the
neighboring vertex $v_{023}$. Two of the sources, $f_0$ and $f_2$ are
already known, only the height $z_3$ of $f_3$ has to be found. First
one has to find the \emph{new} locations of $q_0$ and $q_2$ when
defined from vertex $v_{023}^*$. In order to do this, lines
$\lambda^{023}_i$ from vertex $v_{023}^*$ are drawn, and $q_0$ and $q_2$ are
located using the fact that the heights $z_0$ and $z_2$ are already
known.  Equation (\ref{eq:zi}) implies
\begin{mathletters}
\begin{eqnarray}
 | P_0 q_0^{(023)} |  =   | P_0 q_0^{(012)} |  {  | v^*_{(012)} P_0 | \over  | v^*_{(023)} P_0 | } \\
 | P_2 q_2^{(023)} |  =   | P_2 q_2^{(012)} |  {  | v^*_{(012)} P_2 | \over  | v^*_{(023)} P_2 | }
\end{eqnarray}
\end{mathletters}
Once the new $q_0$ and $q_2$ are known, the next step is to draw
circles $\omega_{03}$ and $\omega_{23}$ through them and normal to the
respective interfaces. Their intersection  gives the location of $q_3$
(this intersection always occurs on $\lambda_3^{023}$, as Theorem \ref{th:recircles}
shows). This determines  $z_3$ and $a_3$ using
equations (\ref{eq:zi}) and (\ref{eq:ai}). This procedure of
triangulation is repeated until all sources are located. Eventually it
may happen that a point $q_i$ is found to be closer to the conjugate
vertex $v^*$ than the corresponding $P_i$. This is not acceptable and
means that the \emph{tentative} starting position of  $q_0$ has to be
changed.  It  must be shifted away from $v_{012}^*$ and the whole
procedure repeated. It is easy to see that taking a large
enough value of $q_0$ always solves this problem.
\\
One could ask whether this construction can be closed
self-consistently.  For example after determining $f_5$ from $f_0$ and
$f_4$ in our example of Fig.~\ref{fig:bubblesources}, one can go on
with the procedure as if $f_1$ were unknown. Would the position of
$f_1$ determined by $f_0$ and $f_5$ be the same one as found initially?
Alternatively: if we used $f_0$ and $f_1$ to determine $f_5$ instead of
going around the bubble in the opposite sense, would its position be
the same as found after going around the bubble? The answer is yes
because of unicity. As discussed in Appendix~\ref{sec:RPrecognition},
the height of one of the sources of a vertex determines the other two
uniquely. This means that $f_1, f_2,\cdots,f_5$ are all uniquely
determined by $f_0$.
\section{Equilibrated Foams.}
\label{sec:equilibrium}
In this section we show that a two dimensional foam in equilibrium
satisfies all conditions required for it to be a SMVP. In order to do
this let us first write down the equilibrium conditions for an arbitrary
vertex of the foam, in compact form. We will consider the case of foams
with arbitrary surface tensions, and  also allow forces to act on the
foam's vertices.
\subsection{Equilibrium equations}
\label{sec:eq_equations}
\begin{figure}[htpb] \vbox{ 
\centerline{\psfig{figure=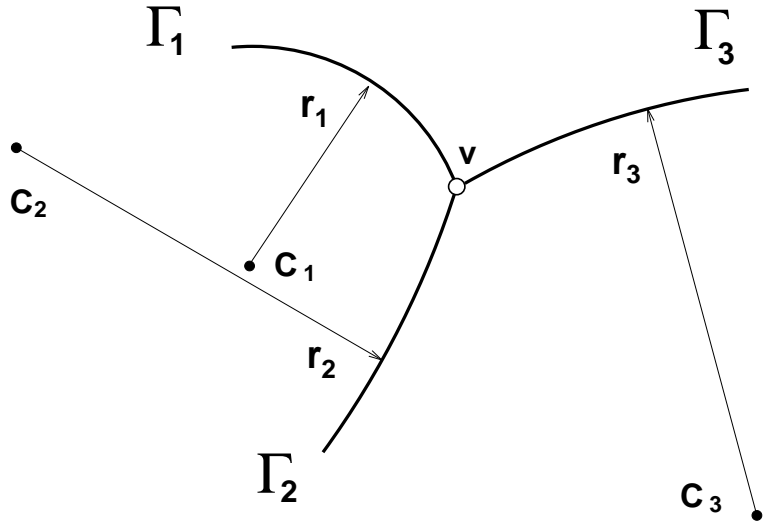,width=10cm,angle=270} }
\centerline{}
\caption{
A generic vertex $v$, at which three interfaces $\Gamma_1$, $\Gamma_2$
and $\Gamma_3$ meet. We regard pressure drops to be positive if the
pressure decreases when the interface is crossed in the counterclockwise
sense of rotation around $v$. Therefore in the case shown in this
figure, $r_2$ and $r_3$ are positive, while $r_1$ is negative.
}
\label{fig:eq_sign_convention}
} \end{figure}

Let $\vec v$ be the location of a vertex $v$ at which three interfaces
$\Gamma_1$, $\Gamma_2$ and $\Gamma_3$ meet, as shown in
Fig.~\ref{fig:eq_sign_convention}. Each interface $\Gamma_i$ is a circle arc with
radius $r_i$ and center $\vec C_i$.  It produces on $v$, due to its
surface tension $\tau_i$, a force $\vec T_i$ of modulus $ 2\tau_i$ in
the direction of the tangent to the film at $\vec v$. These forces $\vec
T_i$ can therefore be written as
\begin{equation}
\vec T_i = - 2 { \tau_i \over r_i }  \vec K_i^* = - \xi_i \vec K_i^* ,
\label{filmforce}
\end{equation}
where $\vec K_i =  \vec C_i -\vec v $ and $\xi_i=2\tau_i/r_i$. The
asterisk stands here for a counterclockwise rotation in $\pi / 2$, so that
${\vec e_x}^*=\vec e_y$.
\\  
Let us more generally assume that an external force $\vec F$ acts on $v$.
Equilibrium of all forces acting on the vertex implies
\begin{equation}
\vec F^* + \sum_{i=1}^3 \xi_i \vec K_i = 0 \qquad \qquad i=1,2,3
\label{forceeq}
\end{equation}
There is still one more condition, which is related to pressure
equilibrium around the vertex. The pressure drop across an interface can
be written as
\begin{equation}
\Delta p_i = 2 { \tau_i \over r_i } = \xi_i
\label{deltap}
\end{equation}
We will adopt the convention for $\Delta p$ to be positive if the
pressure decreases when crossing the interface in the counterclockwise
sense of rotation around $v$. This is of course related to a convention
for the signs of the $r_i$. In Fig.~\ref{fig:eq_sign_convention} $r_1$ and $\Delta
p_1$ are negative according to this convention.  Notice that pressure
jumps and radii have different signs when considered from the two
opposite vertices joined by a film. 
\\
The fact that the total accumulated pressure drop around a vertex has to
be zero implies then
\begin{equation}
\sum_{i=1}^3 \xi_i = 0.
\label{pressureeq}
\end{equation}
\noindent
This is the pressure equilibrium condition for the vertex. 
Equations (\ref{forceeq}) and  (\ref{pressureeq}) are satisfied if the
vertex is equilibrated, and are together equivalent to
\begin{equation}
\vec F^* + \sum_{i=1}^3 \xi_i  ( \vec C_i - \vec x )       = 0,
\label{eq:aligned}
\end{equation}
\noindent
where $\vec x$ is an \emph{arbitrary} point. This is what we will call the equilibrium
condition for the vertex, and encloses both force and pressure equilibrium.
\subsection{Equilibrium implies alignement}
\label{sec:eq_align}
The alignment of the centers of a two-dimensional foam was already known
long time ago by Plateau~\cite{Boys} for equal surface tensions, and in
the case of small self-standing clusters of 2 and 3 bubbles.  For the
case of a cluster of two bubbles and zero load it is a trivial
consequence of symmetry~\cite{Darcy}. The demonstration for the case of
clusters of three bubbles is referred to by Boys as being``rather long
and difficult''~\cite{Boys}.
\\
It is not difficult to see that the alignment of the centers is in no
way a property of clusters, and also not restricted to vertices with
equal surface tensions and zero loads, but a general consequence of
equilibrium.  We will find that under very general conditions such as
arbitrary surface tensions and  external loads applied on the
vertices, if a vertex is equilibrated then the centers of the three
arcs converging to it lay on a line.
\\
Consider equation~(\ref{eq:aligned}), and assume for a moment that the external load is
zero. Then $C_1, C_2$ and $C_3$ lay on the same line, as can be seen by
taking $\vec x = \vec  C_1$ for example. Thus all vertices in
equilibrium are aligned if no external force is applied on it. This
alignment property is even true under non-zero load conditions, if the
force is perpendicular to the line of centers.  A load satisfying such
condition will be called a \emph{normal load}.  The alignment condition
has the geometrical consequence that the interfaces $\Gamma_i$, when
continued, meet each other again at a unique point $v^*$ which we called
the {\it conjugate vertex}~\cite{vertex} of $v$. This also means that
the vertex $v$ could be physically realized as a self-standing cluster
of two bubbles (by continuing its interfaces), and is a kind of a
``separability'' condition for the static equilibrium conditions, in the
sense that each vertex of a foam might as well be that of an isolated
cluster of two bubbles.  Thus
\begin{quote}
A vertex in equilibrium under a normal load is
aligned.
\end{quote}
or ,equivalently
\begin{quote}
A vertex in equilibrium under a normal load has a conjugate vertex.
\end{quote}
Next we would like to consider the existence of a reciprocal figure,
since this is also a condition that has to be satisfied in order for a
foam to be a SMVP.  This condition must be separately considered. The
reader may easily build examples of bubbles with $n$ neighbors, all of
whose $n$ vertices are aligned, but yet do not admit a reciprocal
figure. The reason is that the attempted reciprocal figure will not in
general ``close'' around that bubble, the same case as described in
Appendix~\ref{sec:RPrecognition}.
\subsection{Equilibrium implies Reciprocal Figure}
\label{sec:eq_reciprocal}
\begin{figure}[htpb] \vbox{ 
\centerline{\psfig{figure=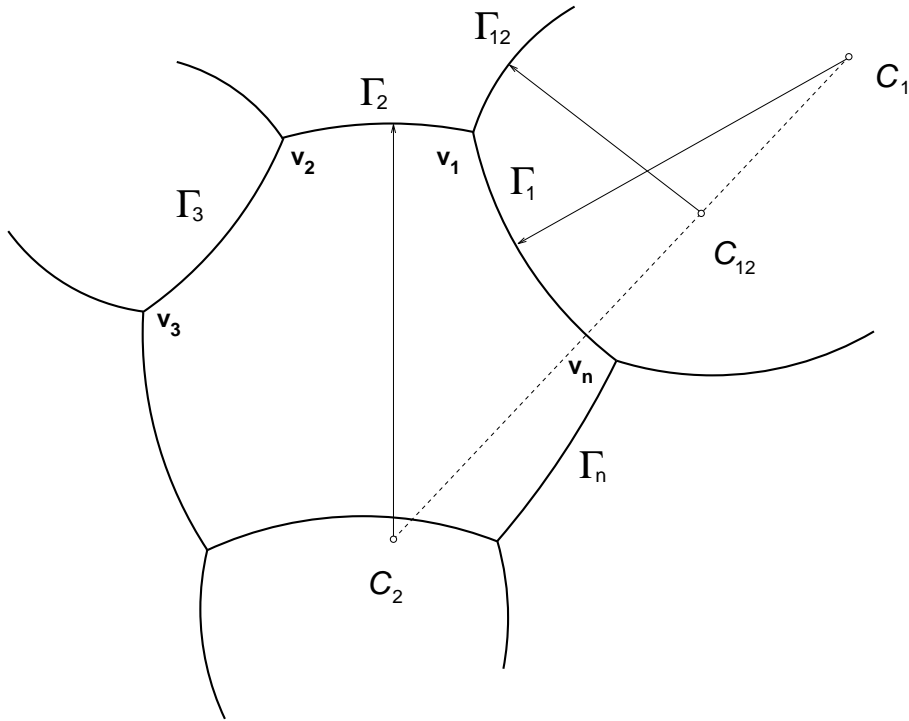,width=12cm,angle=270} }
\centerline{}
\caption{
A bubble surrounded by $n$ neighbors $\alpha=1,\ldots,n$, showing the
naming convention for the interfaces. Notice that film curvatures are of
opposite signs when considered from each of the two vertices at their
ends. Thus for example  $\xi_2$ appears with $+$  sign in the
equilibrium condition of vertex $v_2$, but with $-$ sign in that of
$v_1$ (See equation~\ref{eq:geneq}).
}
\label{fig:bubbleA}
} \end{figure}

\begin{figure}[htpb] \vbox{ 
\centerline{\psfig{figure=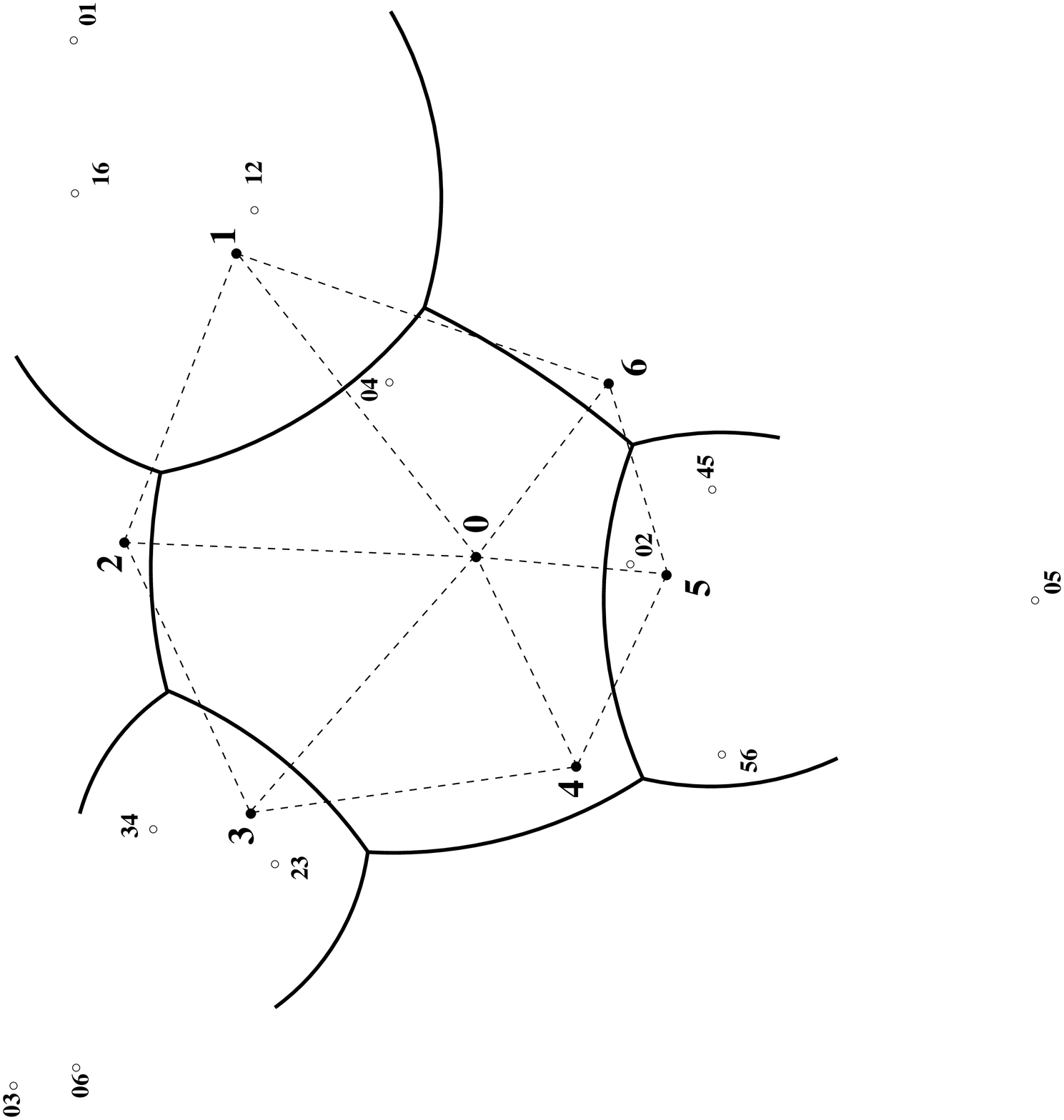,width=12cm,angle=270} }
\centerline{}
\caption{
The same closed bubble of Fig.~\ref{fig:bubbleA}, showing the film
centers $C_{ij}$ (empty dots) aligned in triplets, and the sites
$0,1,\ldots,6$ (black dots) of the reciprocal figure.
}
\label{fig:bubbleC}
} \end{figure}

We will now show that if all vertices of a foam are equilibrated, then
an oriented reciprocal figure exists for it. We start by considering a
bubble and show that the reciprocal figure can be found for it. The
figure for the whole foam can then be formed by patching together those
of neighboring bubbles.  Consider a bubble with $n$ neighbors $\alpha =
1,2,\cdots,n$, as shown in figure \ref{fig:bubbleA}.   At each vertex
$v_\alpha$ of this bubble, three films $\Gamma_\alpha$,
$\Gamma_{\alpha+1}$ and $\Gamma_\alpha^{\alpha+1}$ meet. Interface
$\Gamma_\alpha$ separates the central bubble from its neighbor $\alpha$,
while interface $\Gamma_\alpha^{\alpha+1}$ is the limit between
neighbors $\alpha$ and $(\alpha + 1)$. We will assume vertices
$v_\alpha$ to be in equilibrium under arbitrary \emph{normal} loads
$\vec F_\alpha$. These loads we can generally write as
\begin{equation}
\vec F_\alpha^* = 
-R_{\alpha+1} \vec C_{\alpha+1}
+S_{\alpha}     \vec C_\alpha
+T_\alpha       \vec  C_\alpha^{\alpha+1}
\label{eq:loadgeneral}
\end {equation}
That is, we have decomposed the external load of vertex $v_\alpha$ in
the \emph{dependent} basis formed by the three centers of the films
meeting at the vertex. Because these centers are aligned, and the load
is normal to the line of centers, the following condition is always
satisfied by the coefficients.
\begin{equation}
-R_{\alpha+1} +S_\alpha + T_\alpha = 0.
\label{eq:nullsum}
\end {equation}
The equilibrium condition (\ref{eq:aligned}) now reads
\begin{equation}
- (\xi_{\alpha+1}  + R_{\alpha+1})    (\vec C_{\alpha+1}  - \vec x_\alpha) 
+ (\xi_\alpha         + S_\alpha)          (\vec C_{\alpha  }  - \vec x_\alpha) 
+ (\xi_\alpha^{\alpha+1}+ T_\alpha) (\vec C_\alpha^{\alpha+1} - \vec x_\alpha) =0, 
\qquad\alpha=1, \cdots n
\label{eq:geneq}
\end{equation}
and holds for \emph{arbitrary} $x_\alpha$'s. The sign of $\xi_\alpha$
is determined by the sign convention at vertex $v_\alpha$, therefore
$\xi_{\alpha+1}$ must appear with a minus sign in the equilibrium
equation for vertex $\alpha$.
\\
Since we use a dependent basis (equation.~(\ref{eq:loadgeneral}))  the
coefficients $\{R_{\alpha+1},S_\alpha,T_\alpha \}$ (the
``representation'' of the load) are not uniquely determined for a given
load. There is instead a monoparametric family of coefficients
$\{R_{\alpha+1},S_\alpha,T_\alpha \}$, all satisfying both
(\ref{eq:loadgeneral}) and (\ref{eq:nullsum}), for each load
$F_\alpha$. We will use these degrees of freedom to choose a
representation $\{\tilde R_{\alpha+1},\tilde S_\alpha,\tilde
T_\alpha\}$ that satisfies
\begin{equation}
\tilde S_\alpha = \tilde R_\alpha \qquad \alpha =1,2,\cdots,n
\label{eq:ring}
\end{equation}
Notice that this condition relates the coefficients of two consecutive
loads. We now show that such a representation always exists. We start by
making the degree of freedom in the representation of the loads
explicit. For arbitrary $\nu_\alpha$, we add the null vector
(\ref{eq:geneq}) multiplied by $\nu_\alpha$ to the load
(\ref{eq:loadgeneral}) and get
\begin{equation}
\vec F_\alpha^* = 
-\tilde R_{\alpha+1} \vec C_{\alpha+1}
+\tilde S_{\alpha}     \vec C_\alpha
+\tilde T_\alpha       \vec  C_\alpha^{\alpha+1}	
\label{eq:loadtilde}
\end {equation}
where 
\begin{mathletters}
\begin{eqnarray}
\tilde R_{\alpha+1} & = & R_{\alpha+1} + \nu_\alpha ( R_{\alpha+1} +
\xi_{\alpha+1} )
\label{eq:Rtilde}	\\
\tilde S_{\alpha} & = & S_{\alpha} + \nu_\alpha ( R_{\alpha} +
\xi_{\alpha} )
\label{eq:Stilde}	\\
\tilde T_{\alpha} & = & T_{\alpha} + \nu_\alpha ( T_{\alpha} +
\xi_{\alpha}^{\alpha+1} )
\label{eq:Ttilde}	
\end{eqnarray}
\end{mathletters}
\noindent
Condition (\ref{eq:ring}) then implies 
\begin{equation}
   \nu_{\alpha+1} = { R_{\alpha} - S_{\alpha+1} \over S_{\alpha+1} +
   \xi_{\alpha+1} } + \nu_\alpha           { R_{\alpha} + \xi_{\alpha+1}
\over S_{\alpha+1} + \xi_{\alpha+1} } ,
\end{equation}
which has always a solution in the generic case. 
\\
Using this representation of the loads we can rewrite the
equilibrium condition (\ref{eq:geneq}) as
\begin{equation}
- Q_{\alpha+1} (\vec C_{\alpha+1}  - \vec x_\alpha) 
+ Q_\alpha        (\vec C_{\alpha  }  - \vec x_\alpha) 
+ E_\alpha         (\vec C_\alpha^{\alpha+1}- \vec x_\alpha) =0,
\qquad\alpha=1, \ldots, n
\label{eq:geneq1}
\end{equation}
\noindent
where 
\begin{mathletters}
\begin{eqnarray}
Q_\alpha & = &\xi_\alpha + \tilde  R_\alpha
\label{eq:defQ} \\
E_\alpha  & = & \xi_\alpha^{\alpha+1} + \tilde T_\alpha
\label{eq:defE}
\end{eqnarray}
\end{mathletters}
This amounts to hiding the loads in a redefinition of the surface
tensions and pressures: $\xi_\alpha \to Q_\alpha$. As we have shown,
this can always be done for aligned loads. We notice that the
coefficients
$E_\alpha$ satisfy
\begin{equation}
\sum_{\alpha=1}^n E_\alpha =0 ,
\label{eq:nullaround}
\end{equation}
\noindent
as can be verified by subtracting Eq.~(\ref{eq:geneq1}) with two
different values of $\vec x_\alpha$, and adding up the result for
$\alpha=1,\ldots,n$. 
\\
More generally the fact that all vertices of the bubble are in
equilibrium has the consequence that
\begin{equation}
\sum_{\alpha=1}^n E_\alpha ((\vec C_\alpha^{\alpha+1}- \vec x_0) =0
 \label{eq:vectornullaround}
\end{equation}
for $\vec x_0$ arbitrary. Condition (\ref{eq:vectornullaround}) can be
generalized to any closed path in the foam. The sum is in that case
over all films cut by the closed path. The smallest such  path,
enclosing a vertex, gives the vertex equilibrium condition
(\ref{eq:aligned}). Now that we have written the bubble equilibrium
condition in the convenient form (\ref{eq:geneq1}), we will write 
an algebraic condition for the existence of a reciprocal figure.
\\
The reciprocal figure was defined to be a set of points $\{P_0,
P_1,\cdots, P_n\}$, each associated to a bubble (but not necessarily
contained in it), such that the straight line passing through $ P_i$
and $ P_j$ is normal to the interface $\Gamma_{ij}$ between bubbles $i$
and $j$. This means that $ C_{ij}$, $ P_i$ and $ P_j$ are on the same
line, as in Fig.~\ref{fig:bubbleC}. We will require that $P_0$ and
$P_1$ be arbitrary (with the only condition that $P_0$, $P_1$ are $C_1
$ are on a line). This will allow us to patch together the reciprocal
figures of neighboring bubbles to form that of the whole foam. This is
equivalent to the translation and dilatation degrees of freedom
existent in the definition of a reciprocal figure for a SVP. For
arbitrary $P_0$, take $P_1$ anywhere on the line $(P_0 C_1)$. All other
points are now uniquely determined.  $P_2$ is located in the
intersection of $(P_0 C_2)$ and $(P_1 C_{12})$, next $P_3$ is found as
the intersection of $(P_0 C_3)$ and $(P_2 C_{23})$, etc. When the
figure is closed with the last point $P_n$, there is an extra
condition, since it has to be  the intersection of \emph{three} lines:
$(P_0 C_n)$, $(P_{n-1} C_{n-1 n})$, $(P_1 C_{1n})$.  The construction
of the reciprocal figure is thus an \emph{overdetermined} problem, and
would not in general be possible if the centers $C_{ij}$  were arbitrarily
located.  Let us now write the conditions for this reciprocal figure to
close, in an algebraic form.  The points $P_i$ of the reciprocal figure
are determined by a set of coefficients $\{A_\alpha, B_\alpha\}$
satisfying the following conditions.
\begin{mathletters}
\begin{eqnarray}
(\vec P_\alpha - \vec P_0 ) & = & A_\alpha ( \vec C_\alpha  - \vec P_0)
\label{rfcond1} \\
(\vec P_{\alpha+1} - \vec P_\alpha ) & = & B_\alpha ( \vec
C_\alpha^{\alpha+1} - \vec P_\alpha)
\label{rfcond2}
\end{eqnarray}
\end{mathletters} 
Substituting \ref{rfcond1} into \ref{rfcond2} we get
\begin{equation}
- A_{\alpha+1}  (\vec C_{\alpha+1}  - \vec P_0) 
+ A_\alpha (1 - B_\alpha) (\vec C_\alpha  - \vec P_0) 
+ B_\alpha  (\vec C_\alpha^{\alpha+1} - \vec P_0)  =0
\label{eq:rfcond3}
\end{equation}
\noindent
We will now show that the equilibrium conditions (\ref{eq:geneq1})
ensure that (\ref{eq:rfcond3}) always has a solution, and therefore that
a reciprocal figure exists. Comparison with equation (\ref{eq:geneq1})
lets us conclude that a solution will exist for $P_0$ arbitrary, if we
are able to find a set of coefficients $\{A_\alpha, B_\alpha\}$ that
satisfy
\begin{equation}
{Q_{\alpha+1}   \over  A_{\alpha +1} }	=
{Q_\alpha         \over  A_\alpha(1-B_\alpha) } =
{ E_\alpha        \over B_\alpha }= 
\mu_\alpha, \qquad   \alpha=1, \cdots n
\label{coefcond} 
\end{equation}
\noindent
This is equivalent to
\begin{mathletters}
\begin{eqnarray}
Q_{\alpha+1}  & = & \mu_\alpha A_{\alpha +1}
\label{eq:rel1} \\
Q_\alpha        & = & \mu_\alpha  A_\alpha (1-B_\alpha)
\label{eq:rel2}	\\
E_\alpha         & = & \mu_\alpha  B_\alpha
\label{eq:rel3}
\end{eqnarray}
\end{mathletters}
\noindent
It is not difficult to see that these equations are satisfied if the
$\mu_\alpha$  are related by
\begin{equation}
\mu_{\alpha+1} =  \mu_\alpha + E_\alpha 
\label{eq:recurrence}
\end{equation}
\noindent
Starting from an \emph{arbitrary} $\mu_1$, this recursion relation gives
us the following values of $\mu$. Once all are known,
equation~(\ref{eq:rel1}) provides the values of the $A_\alpha$, which in
turn determine $\{P_1, \cdots, P_n\}$ using (\ref{rfcond1}). The
condition that the figure can be closed is $ P_{\alpha+n}=  P_{\alpha}$,
 and is equivalent to $\mu_{\alpha+n}=\mu_\alpha$.  This condition is
satisfied because  (\ref{eq:nullaround}) holds, and therefore is a
consequence of equilibrium. 
\\
If we were given an arbitrary circular partition, it would not in
general be possible to find a reciprocal figure for it. The fact that
this CP is an equilibrated foam imposes geometrical constraints on it
ensuring, for example, that it admits a reciprocal figure.
\\
We have thus shown that, for an arbitrary equilibrated bubble, it is
always possible to find a reciprocal figure.  We can arbitrarily fix
$P_0$ since $x_\alpha$ in (\ref{eq:geneq1}) can be arbitrarily chosen,
and we can also choose the ``scale'' $|P_0 P_1|$ of the drawing at will,
since the starting value $\mu_1$, that fixes this scale, is arbitrary.
Therefore the reciprocal figures of neighboring bubbles can be patched
together to form a reciprocal figure for the whole foam. Each selection
of a starting point $P_0$ and a ``scale'' $|P_0 P_1|$ determines the
other points uniquely, therefore there are three degrees of freedom in
the determination of the reciprocal figure. Once $P_0$ and $P_1$ are
chosen, all other points are found as intersections of two lines passing
through the centers $C_{ij}$ and one already existing point $P_i$.
\subsection{The orientation condition}
\label{sec:orientation}
\begin{figure}[htpb] \vbox{ 
\centerline{\psfig{figure=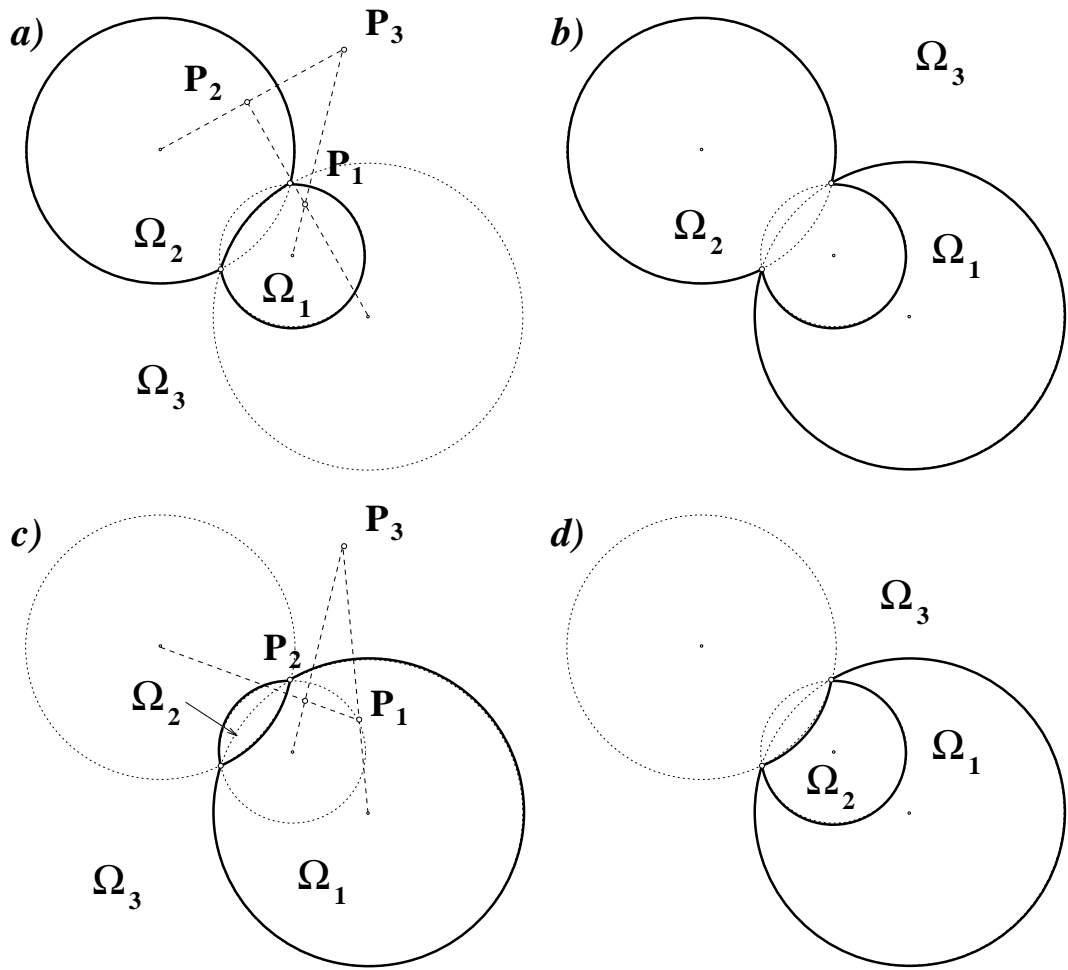,width=14cm,angle=270} }
\centerline{}
\caption{
All these four circular vertices admit a reciprocal figure.  Two of the
vertices, a) and c), are convex and therefore a reciprocal figure
satisfying orientation is possible. The other two are not convex so no
reciprocal figure can satisfy orientation on them.
}
\label{fig:convexity}
} \end{figure}

Now we have to show that it is always possible to find, among all
possible reciprocal figures, at least one that satisfies orientation as
defined in Section~\ref{sec:smvp}. In the first place, if all the
surface tensions are positive then all films will be under traction and
therefore the vertices will be convex. If there are non-convex vertices
in the foam (which would happen if some of the films are compressed
instead of stretched) then we know that it is not possible to satisfy
orientation (Fig.~\ref{fig:convexity}). Positive surface tensions is
thus a necessary condition~\cite{unstable} for the foam to be a SMVP,
although their modulus can be arbitrary for each film.
\\
The orientation condition could fail in the first place because a center
$C_{ij}$ lays in between two points $P_i$ and $P_j$. It is always
possible to avoid this by choosing $P_1$ close enough to $P_0$. In this
way all following points $P_i$ are confined within a (arbitrarily
chosen) small region of space where no center is located. This ensures
that no center $C_{ij}$ lays between $P_i$ and $P_j$.
\\
Now regarding the second part of the orientation condition
(Sec.~\ref{sec:smvp}), consider a vertex $v_{123}$, which is convex and
aligned.  Two orientations of the triangle $P_1 P_2 P_3$ are possible,
as shown in figures \ref{fig:convexity}a and \ref{fig:convexity}c.
Notice that both constitute reciprocal figures for both vertices, but
only one of them satisfies orientation in each case.   In our
construction of the reciprocal figure for the whole foam, we can  decide
the orientation of the initial triangle, choosing the one that respects
orientation. The question is now if the correct orientation of this
starting triangle ensures that of all subsequent ones, whose locations
are determined by $P_0$ and $P_1$. To demonstrate that this is indeed
the case, we notice that the triangle of figure \ref{fig:convexity}a, if
considered as a reciprocal figure for the vertex of
\ref{fig:convexity}c, has all three segments $\overline{P_i P_j}$
wrongly oriented. The point we want to make is that, if the vertex is
convex, there are only two possibilities: either all pairs $P_i P_j$
satisfy orientation, or all are wrongly oriented.  Then if one of the
pairs of a triangle forming part of a reciprocal figure is correctly
oriented, the other two must necessarily also be.  This demonstrates
that if the starting pair $P_0 P_1$ is chosen with the correct
orientation, then all subsequent triangles must be correctly oriented
since they share at least a pair of sources with one preexisting
triangle. Therefore in order to ensure correct orientation of the
whole figure it is enough to correctly choose the orientation of
the first pair. 
\\
\section{Discussion}
\label{sec:conclusions}
\begin{figure}[htpb] \vbox{ 
\centerline{\psfig{figure=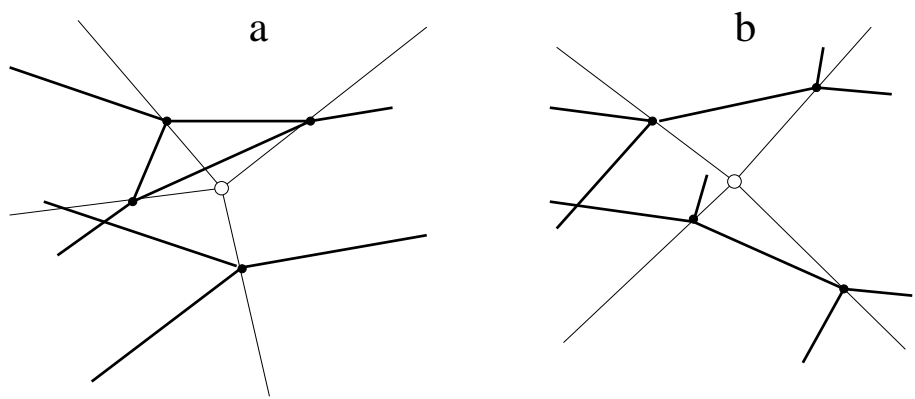,width=12cm,angle=270} }
\centerline{}
\caption{
A three-dimensional Multiplicative \Vo Partition gives rise to vertices
which  simultaneously belong to four neighboring bubbles in a generic
situation. Six spherical interfaces converge at these vertices. When one
such vertex (white dot) crosses the projection plane $\Pi_z$, depending
on its orientation this can  result either in  {\bf a)} a $T2$ process
(disappearance of a triangular bubble) or {\bf b)} a $T1$ process
(neighbor switching) for the two-dimensional foam.
}
\label{fig:process}
} \end{figure}

A dissection of space into cells separated by circular interfaces that
meet at triple vertices is called a Circular Partition (CP). A
two-dimensional foam therefore defines a circular partition of
two-dimensional space.  The equilibrium conditions for the foam impose
geometrical constraints on this CP.  We have here shown that the CP
defined by an equilibrated foam is \emph{aligned} and admits an
\emph{oriented reciprocal figure}.  This result is valid in general for
heterogeneous foams, each of whose films may have an arbitrary
(positive) surface tension, and even if loads are applied on the
vertices, with the sole requirement of equilibrium.  We have  seen in
Sec.~\ref{sec:find_sources}, that any CP satisfying the conditions of
alignement and existence of oriented reciprocal figure is a Sectional
Multiplicative \Vo Partition (SMVP).   A SMVP is a plane cut of a
Multiplicative partition, thus two-dimensional foams are plane cuts of
three-dimensional foams, these last being a multiplicative partition
with respect to sources in three dimensions.  Therefore given an
arbitrary equilibrated two-dimensional foam, it is always possible to
find sources $\{f_i\}$ in three-dimensional space, and amplitudes $a_i$
such that the given foam is a SMVP with respect to those sources.
\\
A first implication of this correspondence is the identification of a
new set of  degrees of freedom for the foam: the intensities $a_i$ and
locations of the sources $f_i$ in  three dimensions. This allows a more natural
description of a foam, than the one that is done in terms of films and
vertices.  When a foam is interpreted as a tessellation of space with
respect to some sources, we see the foam's films and vertices are
secondary constructions, and their evolution is a consequence of that of
the sources. The dynamical description is  conceptually simpler using
the SMVP interpretation. For example the processes of neighbor switching
($T1$) and cell disappearance ($T2$) are described in a unified manner
(Fig.~\ref{fig:process}).  Both are due to the fact that a fourfold
vertex in 3d crosses the projection plane $\Pi_z$.  Depending on the
spatial orientation of the vertex with respect to $\Pi_z$, this is seen
as a $T1$  or $T2$ process.
\\ 
An evolving foam can be seen as a particular instance of a Dynamical Random
Lattice\cite{LHM}, in which the evolution of a cellular structure is
fixed by assigning a given dynamics to the sources of a mathematically
defined tessellation. In the case of foams the dynamics is usually fixed
by gas diffusion across the membranes. Alternatively other dynamical
evolution rules may also be interesting, but in any case one has to make
the translation to obtain the source's dynamics.  The next  step is then
to find, for a given proposed dynamic evolution for the foam, the
corresponding dynamics\cite{CFMtbp} for the sources and intensities.
\\
Foams are usually studied inside a bounded region or cage, which imposes
the only constraint that the films be normal to its boundary.
Boundaries of this kind do not affect the fact that the foam is a SMVP.
We have shown that the result holds with the sole requirement of vertex
equilibrium.  But it is obvious that SMVPs are always closed on
themselves forming a self standing cluster, that is, there are no open
film ends.  This implies that even bounded foams \emph{must} be a region
of a larger self-standing cluster of bubbles that is closed on itself,
and everywhere equilibrated. The point is not trivial in that it ensures
that all films ending at the boundary can be continued, eventually
forming new (phantom) vertices, and that the resulting foam will have
all its vertices in equilibrium. We see then that there is no
fundamental difference between bounded and self-standing foams, since
all foams are regions of a self-standing cluster. This does not mean
that the boundaries have no effect, which would of course be false. If
the foam's dynamic inside the cage is for example produced by gas
diffusion across the films, the evolution of the ``phantom'' bubbles
existing as continuations of the physical foam outside the cage, will
\emph{not} follow this dynamic, but a different one, which is determined
by the constraint that the films be normal to the cage's boundaries.
\\
In the field of joint-bar structures, an old result due to
Maxwell\cite{Maxwell1,Maxwell2} states that if a lattice accepts a
reciprocal figure then it can support a self-stress, and conversely.
More recently Ash and Bolker\cite{AshBolker2} have shown that the
existence of a reciprocal figure is sufficient condition for the lattice
to be a Sectional \Vo Partition. In this case there is the additional
requirement that all vertices are convex, therefore all stresses in the
lattice must be of the same sign, and the lattice can be an equilibrated
spider web. A chain of results that span a century allow us to see
equilibrated spider webs as SVPs , and conversely.  The alert reader may
have noticed that equilibrated foams can be seen as a kind of
``generalized'' spider webs, in which the pressure difference between
cells is the new ingredient, and is equilibrated by the curvature of the
interfaces.  It therefore turns out as no surprise that these
generalized spider webs (foams) are equivalent to an appropriate
generalization of SVPs, namely SMVPs, which include a multiplicative
constant that gives rise to curved interfaces.
\\
\acknowledgments
I have benefited from discussions with K.~Lauritsen, H.~Herrmann,
H.~Flyvbjerg, D.~Le Caer and D.~Weaire. I thank D.~Weaire for reference
\cite{Boys}, and D.~Le Caer for making me aware of Plateau's results and
sending me references \cite{Boys} and \cite{Darcy}. I am also grateful
to H.~Flyvbjerg and D.~Stauffer for their efforts to help me make this
work more readable.
\\
\vfill \eject
\appendix
\section{The recognition problem for rectilinear partitions}
\label{sec:RPrecognition}
\begin{figure}[htpb] \vbox{ 
\centerline{\psfig{figure=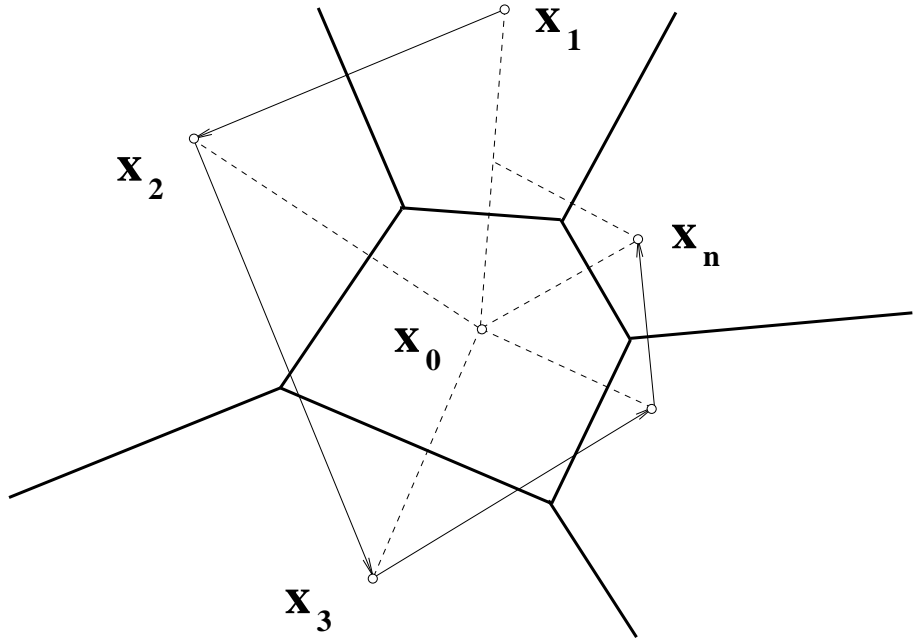,width=12cm,angle=270} }
\centerline{}
\caption{
An arbitrary rectilinear partition will not in general admit a
reciprocal figure. The one in this example does not admit a reciprocal
figure, and therefore cannot be a Sectional \Vo Partition.
}
\label{fig:closing}
} \end{figure}

\begin{figure}[htpb] \vbox{ 
\centerline{\psfig{figure=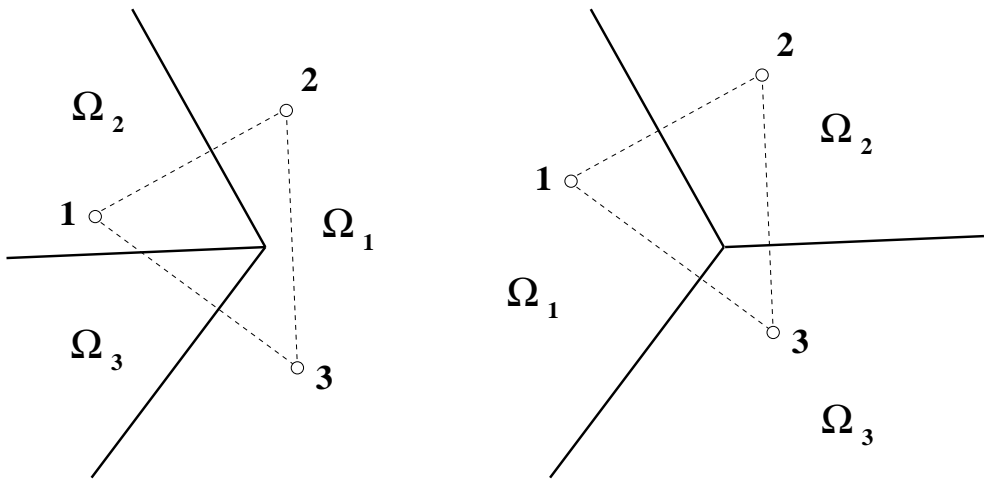,width=13cm,angle=270} }
\centerline{}
\caption{
Both these partitions with three cells admit a reciprocal figure. The
one on the left does not satisfy orientation and therefore cannot be
part of a Sectional \Vo Partition. The one on the right satisfies
orientation, and therefore it is a SVP. It is then possible to find
sources $f_1$, $f_2$ and $f_3$ in three-dimensional space, such that
this partition is obtained as the cut $z=0$ of a three-dimensional \Vo
Partition with respect to $\{f_i\}$. Sources $f_i$ will be located at
heights $z_i$ above points $\{1, 2, 3\}$ in this figure.
}
\label{fig:orientation}
} \end{figure}

A  classification of space into cells $\Omega_i$ separated by
rectilinear interfaces $\Gamma_{ij}$ is called a \emph{rectilinear
partition} of space. Given a rectilinear partition $\cal P$ of the
plane, a {\it reciprocal figure} $\cal R(\cal P)$is a planar graph
composed of sites $i$ joined by edges $(ij)$
satisfying~\cite{Maxwell1,AshBolker2}:
\begin{itemize}
\item sites $i$ of $\cal R$ are in one-to-one correspondence with cells
$\Omega_i$ of $\cal P$.
\item edges $(ij)$ of $\cal R$ are in one-to-one correspondence with 
interfaces $\Gamma_{ij}$ of $\cal P$.
\item edges $(ij)$ of $\cal R$ are \textbf{normal} to interfaces
$\Gamma_{ij}$ of $\cal P$.
\end{itemize}
Clearly a reciprocal figure is defined up to arbitrary global
dilatations and translations, since angles are not changed by them.
Therefore, if $\cal P$ admits a reciprocal figure, there are three
degrees of freedom in its determination~\cite{Maxwell1,AshBolker2} . An
arbitrary partition $\cal P$ will \emph{not} in general admit a
reciprocal figure. We can see this with a simple example. Draw an
arbitrary polygonal cell $\Omega_0$ with $n$ faces, and take arbitrary
rectilinear interfaces between its $n$
neighbors(Fig.~\ref{fig:closing}). Now take an arbitrary point $x_0$ on
the plane to start with, an assign it to the central cell. This starting
point is arbitrary since a reciprocal figure is defined up to arbitrary
translations. The other $n$ points $\{x_1, \dots, x_n\}$ associated to
the external cells must be somewhere on the $n$ rays $r_i$ stemming from
$x_0$ and normal to the faces $\Gamma_{0i}$ of $\Omega_0$.  The global
length scale is also arbitrary so that say $\overline{x_0 x_1}$ can be
freely chosen. Then we choose $x_1$ somewhere on $r_1$.  Now point $x_2$
is determined as the intersection of ray $r_2$ with the normal to face
$\Gamma_{12}$ going through $x_1$.  This can be repeated to obtain all
$n$ external points, but in general the figure will not \emph{close},
that is, segment $\overline{x_n x_1}$ will not be normal to interface
$\Gamma_{1n}$, whose orientation is arbitrary.
\\
The planar graph formed by joining the sources $P_i$ of a SVP with edges
$(ij)$, one for each nonempty interface $\Gamma_{ij}$, constitutes a
reciprocal figure for the SVP. Therefore every SVP has a reciprocal
figure.
$$
\cal P  \hbox{ is a SVP } \Rightarrow \cal P \hbox{ admits a reciprocal figure}
$$
Recently Ash and Bolker have shown that the existence of a reciprocal
figure satisfying \emph{orientation} ~\cite{AshBolker2} is also a
\emph{sufficient} condition for $\cal P$ to be a SVP.  A reciprocal
figure $\cal R$ satisfies orientation if for each bond $ij$, the sites
$i$ and $j$ are oriented in the same way as cells $\Omega_i$ and
$\Omega_j$ are with respect to $\Gamma_{ij}$.
$$
\cal P  \hbox{ admits an oriented reciprocal figure} \Rightarrow \cal P
\hbox{ is a SVP}
$$
This result solves the recognition problem for SVPs. The orientation
condition can only be satisfied if all vertices of $\cal P$ are
\emph{convex}. We will say that a vertex is convex if the internal
angles formed by the interfaces are all smaller than $\pi$.
Figure~\ref{fig:orientation} shows two partitions with three cells.  One
of them is convex, the other  is not. In the second case it is not
possible to find a reciprocal figure that satisfies orientation, and
therefore it cannot be a SVP.
\\
Given a rectilinear partition $\cal P^*$, and an oriented reciprocal
figure for it (as in Fig.~\ref{fig:recfig}), it is always
possible\cite{AshBolker2} to find sources $\{f_i\}$ in three dimensions,
located at heights $z_i$ above the points $P_i$, such that $\cal P^*$ is
the section with $\Pi_z$ of a three-dimensional VP with respect to
$\{f_i\}$.
The procedure to determine the heights $z_i$ can be easily described.
Vertices $v_{ijk}$ of $\cal P^*$ will be equidistant from sources $f_i$,
$f_j$, $f_k$ (see Fig.~\ref{fig:SVPlocate_sources}).  Start from an
arbitrary vertex, say $v_{ijk}$, and draw a spherical surface of
\emph{arbitrary} radius $r_{ijk}=r_0$ with center at that vertex. Now
define sources $f_i$, $f_j$ and $f_k$  as the intersections of this
surface with the verticals (normals to $\Pi_z$) through $P_i$, $P_j$ and
$P_k$ respectively.  Next go to a neighboring vertex, which shares two
sources with this one. Let us call it $v_{ijm}$. For this vertex, only
source $f_m$ has to be located since $f_i$ and $f_j$ are known. Draw a
spherical surface with center $v_{ijm}$ and containing $f_i$ and $f_j$.
Both will be simultaneously contained, since interface $\Gamma_{ij}$, on
which $v_{ijm}$ is located, is equidistant from $f_i$ and $f_j$. The
radius $r_{ijm}$ of this surface is determined by the locations of $f_i$
and $f_j$, which in turn is fixed by $r_0$. Its intersection with the
vertical $n_m$ through $P_m$ determines $f_m$. If this spherical surface
does not intersect $n_m$, just choose a larger value of $r_0$ and start
all over again (from the initial vertex).  The construction proceeds in
this manner until all sources have been determined. As mentioned, the
initial value of $r_0$ is tentative, in the sense that it may have to be
modified (increased) if at some point during the procedure, a normal
line is not cut by the corresponding spherical surface from the vertex.
It is easy to see that increasing the value of the starting radius $r_0$
is always enough to solve this problem.
\\
There is thus one degree of freedom in this construction ($r_0$). We
conclude that, given a two-dimensional partition $\cal P$ that admits a
reciprocal figure, there is a four-parametric family of source locations
such that $\cal P$ is a SVP with respect to them.  Three degrees of
freedom come from the determination of the reciprocal figure itself
(since a dilatation and/or translation of a reciprocal figure is again a
reciprocal figure) and the last one from $r_0$. This last degree of
freedom results from the fact that a SVP is invariant if all heights are
changed according to $z_i^2 \to z_i^2 + \alpha^2$ with $\alpha$
arbitrary (see equations (\ref{eq:svpcell}) and (\ref{eq:svpinterface})).
\\
Reciprocal figures were first studied by
Maxwell~\cite{Maxwell1,Maxwell2} in relation with the rigidity of
bar-joint frameworks in the plane.  Maxwell pointed out that frameworks
that have a reciprocal figure are able to support a self-stress, and
conversely.  The reason is that the edges of the reciprocal figure can
be taken to represent forces transmitted by the edges of the original
framework (rotated by $\pi/2$).  Since these edges form closed polygons,
the existence of a reciprocal figure implies the existence of an
equilibrated set of internal stresses in the absence of external load.
The addition of the orientation condition  (a condition not required by
Maxwell's definition of a reciprocal figure) has the statical
consequence that all signs of the stresses are equal, for example all
traction or all compression. It is clear that no equilibrium is possible
in the case of Fig.~\ref{fig:orientation}a if all three stresses are to
have the same sign.  Fig.~\ref{fig:orientation}b on the other hand, can
be in equilibrium under compression or traction on all three interfaces.
\\
The conclusion is that every SVP is an equilibrium configuration for a
spider web~\cite{AshBolker2}, and conversely, each such equilibrium
configuration is a SVP.
\\
The existence of a reciprocal figure has also projective consequences,
which have been studied by Crapo~\cite{Crapo} and
Whiteley~\cite{Whiteley}.
\\
\section{The Inversion Transformation }
\label{sec:inversion}
We briefly describe here a  geometric transformation called
\emph{inversion}~\cite{Coxeter}.  We will find it extremely useful for
our purpose of discussing circular partitions. An inversion with radius
$k$ around a point $O$ located at $\vec r_0$ transforms a point $P$
located at $\vec r_0 + \vec r $ into a point $P'$ at $\vec r_0 + \vec r\
'$ satisfying
\begin{equation}
\vec r \ '= { k^2  \over r^2 } \quad \vec r
\label{eq:inversion}
\end{equation}
where $r = | \vec r |$.
The sphere of radius $k$ and centered at $O$ is invariant under this
transformation, while the inside and outside of this sphere are
interchanged. Obviously this transformation is self-inverse. Let us now
describe some of the properties of this transformation in two
dimensions~\cite{Coxeter}.  Most of them apply trivially in higher
dimensions.
\begin{itemize}
\item   Circles not through $O$ are transformed in circles not trough
$O$.
\item   Circles through $O$ are transformed into straight lines not
through $O$.
\item Straight lines not through $O$ are transformed in circles through
$O$.
\item Straight lines through $O$ are invariant.  
\item Angles are preserved (in modulus) by the inversion.
\end{itemize}
Given two points $P_1$ and $P_2$ at distances $r_1$ and $r_2$ from the
inversion center, the distance $d_{12}$ between them
transforms as
\begin{equation}
d_{12} '  = d_{12}  { k^2 \over r_1 r_2 }
\label{eq:dist_transf}
\end{equation}
Using this result we can easily see that the inversion is a ``symmetry''
of a MVP in any dimension if the intensities are also appropriately
transformed~\cite{AshBolker2}.  More precisely, a MVP of $R^n$ with
respect to sources $\{P_i\}$ with intensities $\{a_i\}$ is transformed
by an inversion into a MVP with respect to $\{P_i'\}$ with intensities
$\{a_i'\}$, where the new intensities satisfy
\begin{equation}
a_i' =  { a_i \over r_i } \quad A_0
\label{eq:transforma}
\end{equation}
Here $A_0$ is an arbitrary prefactor, the same for all $a_i$'s, and
$r_i$ is the distance between source $i$ and the inversion center $O$.
To see this it suffices to demonstrate that if $x \in \Gamma_{ij}$ then
after an inversion, $x' \in \Gamma_{ij}'$, which is easily done using
(\ref{eq:mvpinterface}), (\ref{eq:dist_transf}) and
(\ref{eq:transforma}).  The inversion transformation is of course also
a symmetry of the SMVP (Section~\ref{sec:smvp}), if the inversion
center $O$ is on the cutting plane $\Pi_z$, since in this case the inversion
leaves this plane invariant.
\\
If the inversion center $O$ happens to be located on an interface
$\Gamma_{ij}$ of a MVP (initially a spherical surface), the transformed
interface $\Gamma_{ij}'$ is a plane not through $O$. The resulting
interface thus corresponds to a \Vo Partition with respect to
sources $i$ and $j$ in their new locations.  Therefore the transformed
intensities $a_i'$ and $a_j'$ have to be equal after the inversion,
which is verified using (\ref{eq:transforma})
\begin{equation}
O \in \Gamma_{ij} \Rightarrow { a_i \over r_i } = { a_j \over r_j }
\Rightarrow a_i' = a_j'
\label{eq:ongamma}
\end{equation}
In the same way a SMVP with respect to two sources $i$ and $j$ is
transformed into a SVP if $O \in \Gamma_{ij}$. The intensities $a_i$
are transformed according to (\ref{eq:transforma}), where $r_i$ is the
distance between $O$ and source $f_i$ in three-dimensional space. The
way in which heights $z_i$ transform is easily found using
(\ref{eq:inversion}). Notice that if the inversion center coincides
with the location of a conjugate vertex $v^*$, then the transformed
partition has a rectilinear vertex since three interfaces are
simultaneously transformed into straight lines. We will use this property of
the inversion transformation when solving the recognition problem for
SMVPs in Section~\ref{sec:find_sources}. 
%%%%%%%%%%%%%%%%%%%%%%%%%%%%%%%%%%%%%%%%%%%%%%
%                                                                   %
%                         REFERENCES                                %
%                                                                   %
%%%%%%%%%%%%%%%%%%%%%%%%%%%%%%%%%%%%%%%%%%%%%%
%

%
\newpage
\centerline{\bf FIGURES}
\bigskip
%%%%%%%%%%%%%%%%%%%%%%%%%%%%%%%%%%%%%%%%%%%%%%%%%
%			FIGURES 
%%%%%%%%%%%%%%%%%%%%%%%%%%%%%%%%%%%%%%%%%%%%%%%%% 
%%%%%%%%%%%%%%%%%%%%%%%%%%%%%%%%%%%%%%%%%
%%%%%%%%%%%%%%%%%%%%%%%%%%%%%%%%%%%%%%%%%
%%%%%%%%%%%%%%%%%%%%%%%%%%%%%%%%%%%%%%%%%
%%%%%%%%%%%%%%%%%%%%%%%%%%%%%%%%%%%%%%%%%
%%%%%%%%%%%%%%%%%%%%%%%%%%%%%%%%%%%%%%%%%
%%%%%%%%%%%%%%%%%%%%%%%%%%%%%%%%%%%%%%%%%
%%%%%%%%%%%%%%%%%%%%%%%%%%%%%%%%%%%%%%%%%
%%%%%%%%%%%%%%%%%%%%%%%%%%%%%%%%%%%%%%%%%
%%%%%%%%%%%%%%%%%%%%%%%%%%%%%%%%%%%%%%%%%
%%%%%%%%%%%%%%%%%%%%%%%%%%%%%%%%%%%%%%%%%
%%%%%%%%%%%%%%%%%%%%%%%%%%%%%%%%%%%%%%%%%
%%%%%%%%%%%%%%%%%%%%%%%%%%%%%%%%%%%%%%%%%
%%%%%%%%%%%%%%%%%%%%%%%%%%%%%%%%%%%%%%%%%
%%%%%%%%%%%%%%%%%%%%%%%%%%%%%%%%%%%%%%%%%
\vbox{
%%%%%%%%%%%%%%%%%%%%%%%%%%%%%%%%%%%%%%%%%
\begin{figure}[htpb] \vbox{ 
\centerline{\psfig{figure=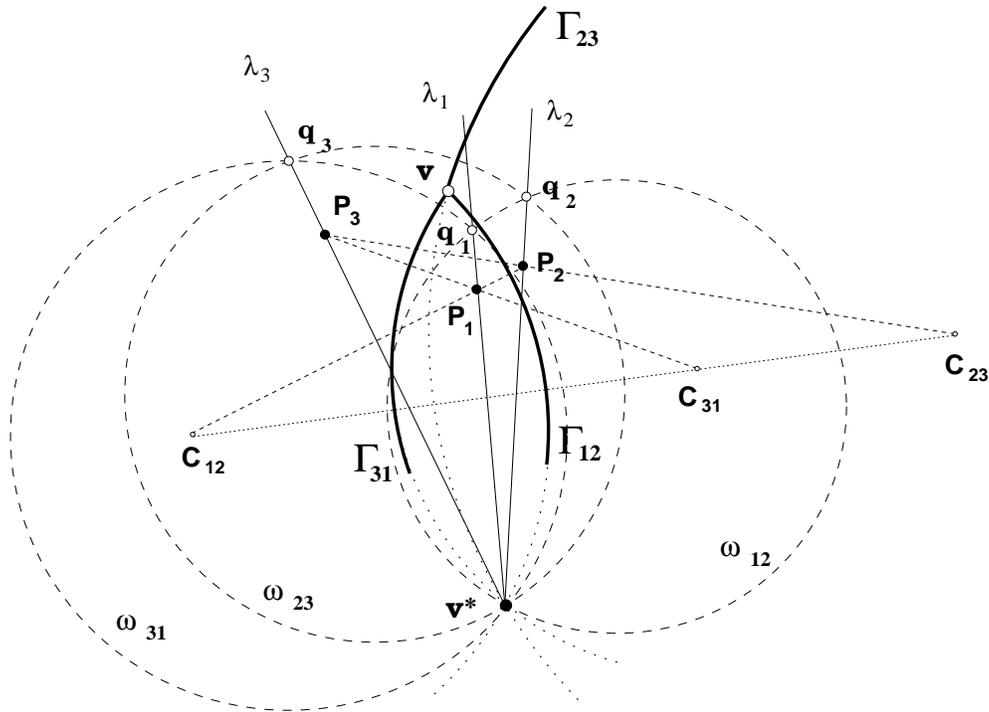,width=13cm,angle=270} }
\centerline{}
\caption{
A vertex $v$ with its reciprocal figure $P_1 P_2 P_3$, showing circles
$\omega_{ij}$. These circles are normal to the interfaces $\Gamma_{ij}$,
and intersect each other at points $q_i$ on $\lambda_i$.
}
\label{fig:reciproca1}
} \end{figure}
%%%%%%%%%%%%%%%%%%%%%%%%%%%%%%%%%%%%%%%%%
%%%%%%%%%%%%%%%%%%%%%%%%%%%%%%%%%%%%%%%%%
\begin{figure}[htpb] \vbox{ 
\centerline{\psfig{figure=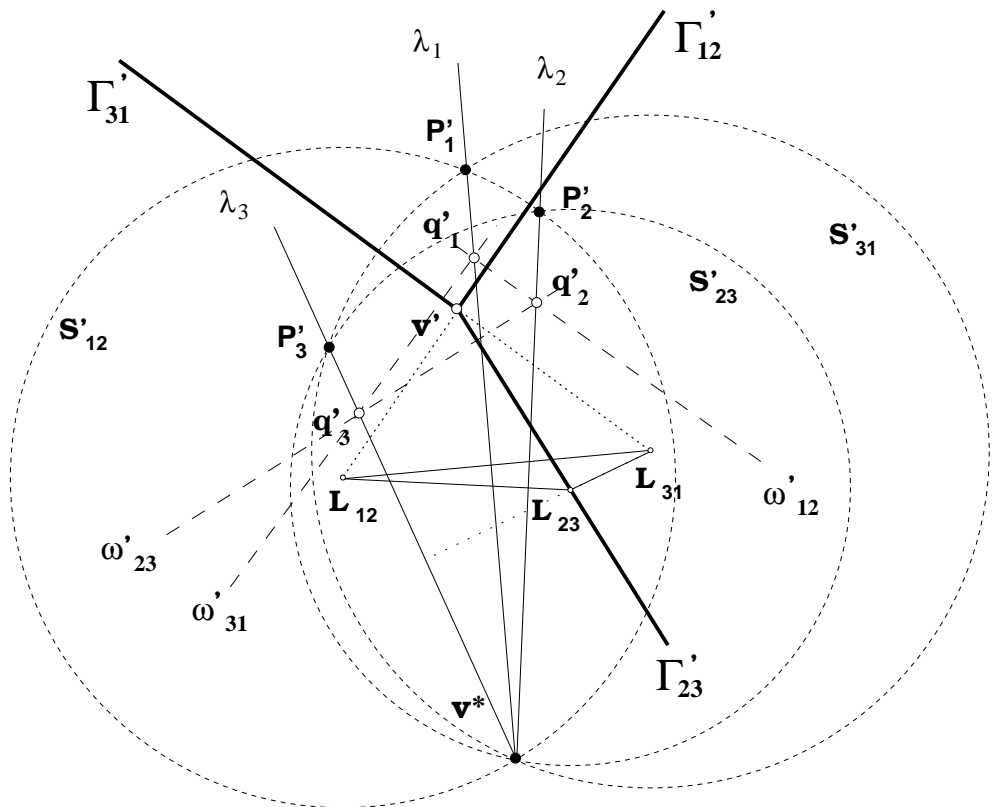,width=13cm,angle=270} }
\centerline{}
\caption{
Same as Fig.~\ref{fig:reciproca1}, after an inversion around $v^*$. All
circles originally through $v^*$ are now straight lines. This is the
case of interfaces $\Gamma_{ij}$ and therefore we call this the
``straight'' representation of the vertex.
}
\label{fig:reciproca2}
} \end{figure}
%%%%%%%%%%%%%%%%%%%%%%%%%%%%%%%%%%%%%%%%%
}
%%%%%%%%%%%%%%%%%%%%%%%%%%%%%%%%%%%%%%%%%
%%%%%%%%%%%%%%%%%%%%%%%%%%%%%%%%%%%%%%%%%
%%%%%%%%%%%%%%%%%%%%%%%%%%%%%%%%%%%%%%%%%
%%%%%%%%%%%%%%%%%%%%%%%%%%%%%%%%%%%%%%%%%
%%%%%%%%%%%%%%%%%%%%%%%%%%%%%%%%%%%%%%%%%
%%%%%%%%%%%%%%%%%%%%%%%%%%%%%%%%%%%%%%%%%
%%%%%%%%%%%%%%%%%%%%%%%%%%%%%%%%%%%%%%%%%
%%%%%%%%%%%%%%%%%%%%%%%%%%%%%%%%%%%%%%%%%
%%%%%%%%%%%%%%%%%%%%%%%%%%%%%%%%%%%%%%%%%
%%%%%%%%%%%%%%%%%%%%%%%%%%%%%%%%%%%%%%%%%
%%%%%%%%%%%%%%%%%%%%%%%%%%%%%%%%%%%%%%%%%
%%%%%%%%%%%%%%%%%%%%%%%%%%%%%%%%%%%%%%%%%
%%%%%%%%%%%%%%%%%%%%%%%%%%%%%%%%%%%%%%%%%
%%%%%%%%%%%%%%%%%%%%%%%%%%%%%%%%%%%%%%%%%
%%%%%%%%%%%%%%%%%%%%%%%%%%%%%%%%%%%%%%%%%
%%%%%%%%%%%%%%%%%%%%%%%%%%%%%%%%%%%%%%%%%
\vbox{
%%%%%%%%%%%%%%%%%%%%%%%%%%%%%%%%%%%%%%%%%
\begin{figure}[htpb] \vbox{ 
\centerline{\psfig{figure=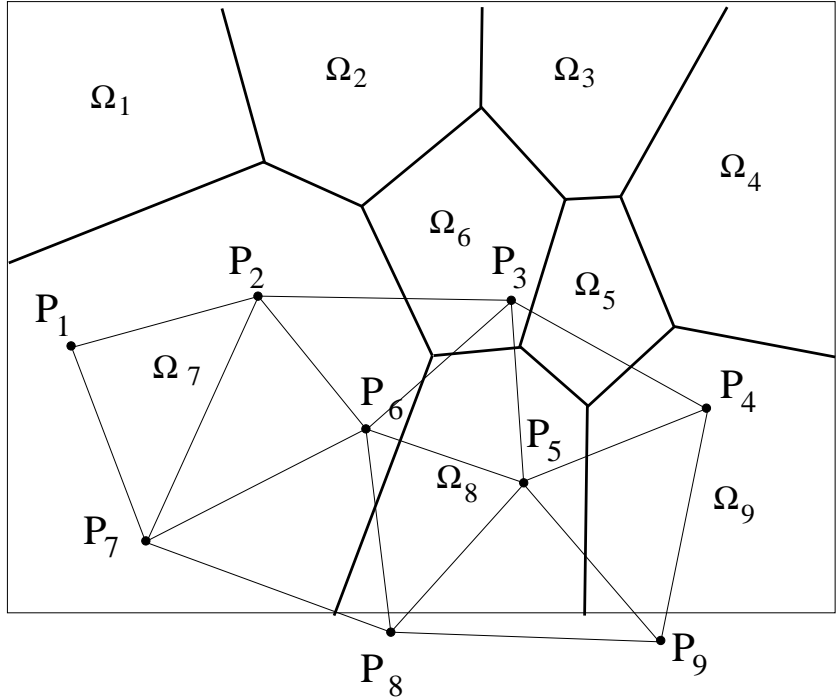,width=11cm,angle=270} }
\centerline{}
\caption{
A convex rectilinear partition (thick lines) with cells
$\{A,B,\ldots,I\}$ and its associated reciprocal figure (thin lines)
with sites $\{a,b,\ldots,i\}$ (black dots). This reciprocal figure (or
any other figure obtained from dilatation and translation of this one)
satisfies orientation, and therefore the partition is a Sectional \Vo
Partition. Sources of this SVP are located above points $P_i$.
}
\label{fig:recfig}
} \end{figure}
%%%%%%%%%%%%%%%%%%%%%%%%%%%%%%%%%%%%%%%%%
%%%%%%%%%%%%%%%%%%%%%%%%%%%%%%%%%%%%%%%%%
\begin{figure}[htpb] \vbox{ 
\centerline{\psfig{figure=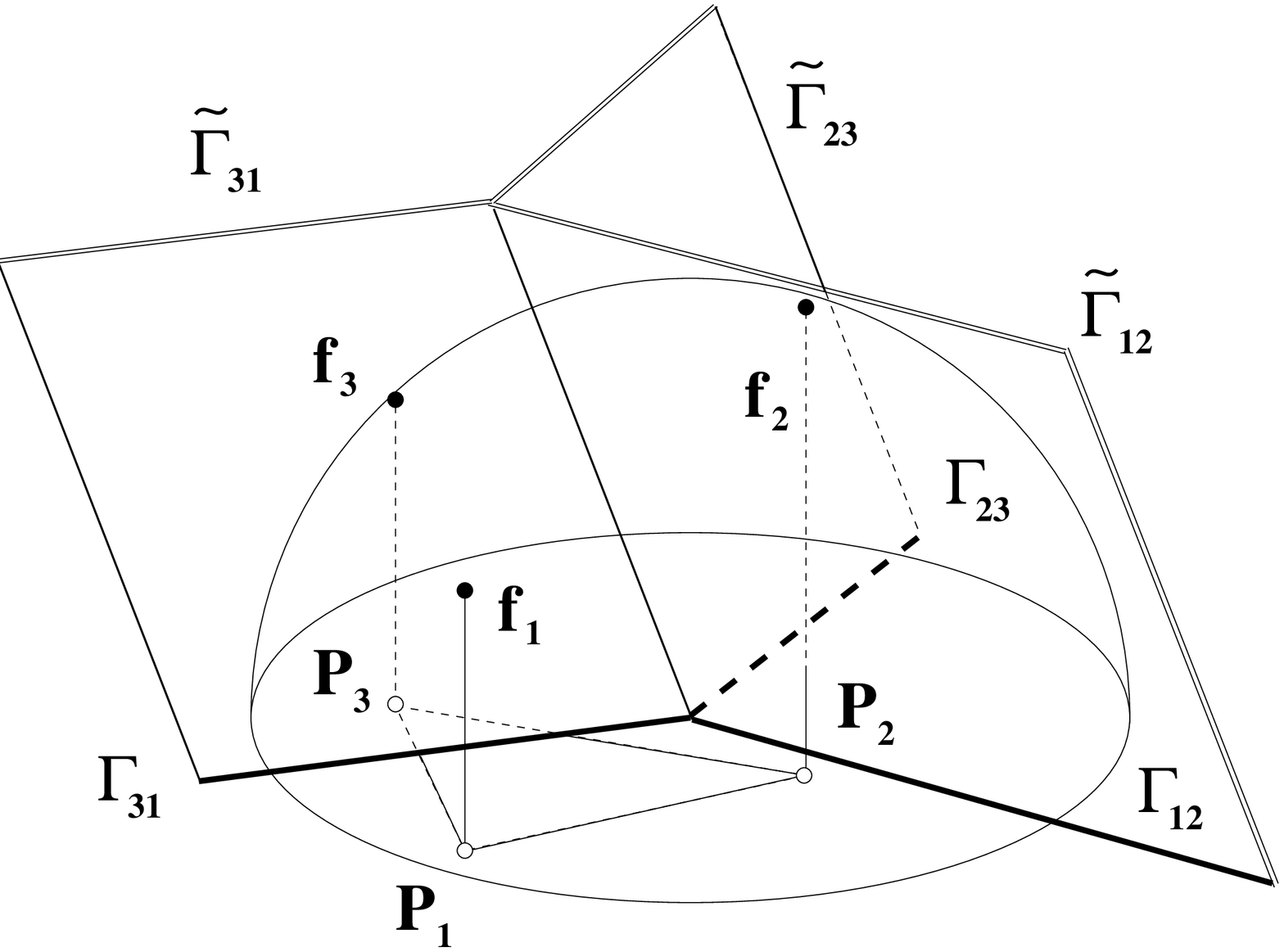,width=11cm,angle=270} }
\centerline{}
\caption{
Given an oriented reciprocal figure $\{P_1, P_2, P3\}$ (white dots) for
a rectilinear partition (interfaces $\Gamma_{ij}$), the sources $f_i$ in
three dimensional space can be located above the points $P_i$. These
sources must be equidistant from the vertex, therefore they are at the
intersections (black dots) of a spherical surface centered at the vertex
$v_{123}$ with the normals through $P_i$.
}
\label{fig:SVPlocate_sources}
} \end{figure}
%%%%%%%%%%%%%%%%%%%%%%%%%%%%%%%%%%%%%%%%%
}

\begin{references}
%
\bibitem{Rev1} N.~Rivier, {\it On the structure of random tissues or
froths, and their evolution.}, Philosophical Magazine B {\bf 47}
(1983), L45-L49.
%
\bibitem{Rev2} D.~Weaire and N.~Rivier, {\it Soap, Cells and
Statistics---Random Patterns in Two Dimensions}, Contemp.~Phys.~{\bf
25} (1984), 59-99.
%
\bibitem{Rev3} J.~Glazier and D.~Weaire, {\it The kinetics of cellular
patterns}, J.Phys.:Condens.~Matter {\bf 4}, (1992), 1867-1894.
%
\bibitem{Rev4} J.~Stavans, {\it The evolution of cellular patterns},
Rep.~Prog.~Phys. {\bf 56} (1993), 733-789.
%
\bibitem{Smith1} C.~S.~Smith, {\it The Shape of Things}, Sci.~Am.~{\bf
190} (1954), 58.
%
\bibitem{Smith2} C.~S.~Smith, {\it Grain shapes and other metallurgical
applications of topology}, in: Metal Interfaces, Am.~Soc. Metals,
Cleveland, 1952.
%
\bibitem{Numerical_Foam}  
D.~Weaire and J.~Kermode, {\it Computer simulation of a two-dimensional
soap froth I. Method and motivation}, Philosophical Magazine B~{\bf 48}
(1983), 245-259.\\
J.~Kermode and D.~Weaire, {\it 2D-FROTH: a program for the
investigation of 2-dimensional froths}, Comput.~Phys.~Commun.~{\bf 60}
(1990), 75-109. \\
T.~Herdtle and H.~Aref, {\it Numerical experiments on a
two-dimensional foam}, J.~Fluid Mech.~{\bf 241} (1992), 233-260. 
%
\bibitem{Flyvbjerg} H.~Flyvbjerg and C.~Jeppesen, {\it A Solvable Model
for Coarsening Soap Froths and Other Domain Boundary Networks in Two
Dimensions}, Physica Scripta {\bf T38}, (1991), 49-54.
\\
H.~Flyvbjerg, {\it Dynamics of Soap Froth}, Physica A {\bf 194},(1993),
298.
\\
H.~Flyvbjerg, {\it Model for coarsening froths and foams}, Phys.~Rev.~E
{\bf 47}, (1993), 4037.
%
\bibitem{Vertex} K.~Kawazaki, T.~Nagai and K.~Nakashima, Phil.~Mag.~B
{\bf 60}, (1989), 399.
\\
K.~Kawasaki, Physica A {\bf 163}, (1990), 59.
%
\bibitem{Boots_book} {\it Spatial Tessellations. Concepts and
Applications of Voronoi Diagrams.}, A.~Okabe, B.~Boots and K.~Sugihara,
Wiley 1992.
%
\bibitem{Honda} H.~Honda, {\it Description of Cellular Patterns by
Dirichlet Domains: The Two-Dimensional Case}, J.~Theor. Biol. {\bf 72},
(1978), 523.
%
\bibitem{Icke} V.~Icke and R. van de Weygaert, {Fragmenting the
Universe. I. Statistics of two-dimensional Voronoi foams}, Astron.
Astrophys. {\bf 184}, (1987), 16-32.
%
\bibitem{Imai} Hiroshi Imai, Masao Iri and Kazuo Murota, {\it \Vo
diagram in the Laguerre geometry and its applications.}, SIAM J.~Comput.
{\bf 14}, (1985), 93.
%
\bibitem{Boots} B.~N.~Boots, {\it Weighting Thiessen Polygons}, {\it
Economic Geography} (1979), 248-259.
%
\bibitem{Boys} {\it Soap Bubbles}, C.~Vernon Boys,  NY: Dover 1959,
pp. 120-127. \\ Reprinted in: {\it The World of Mathematics }, Tempus
Books of Microsoft Press 1988 vol. II, pp. 883-886.
%
\bibitem{Darcy} {\it On Growth and Form}, D'arcy Wentworth Thompson,
Cambridge University Press 1961, page 96.
%
\bibitem{Maxwell1} Maxwell, J.~C. {\it On Reciprocal Figures and Diagrams
of Forces}, {\it Phil.~Mag. Series} {\bf 4} (1864),250-261.
%
\bibitem{Maxwell2} Maxwell, J.~C. {\it On Reciprocal Figures, Frames, and
Diagrams of Forces}, {\it Trans.~Royal Soc.~Edinburgh} {\bf 26}
(1869-72),1-40.
%
\bibitem{AshBolker2} Peter F.~Ash and Ethan D.~Bolker, {\it Generalized
Dirichlet Tessellations}, {\it Geometriae Dedicata} {\bf 20}
(1986),209-243.
%
\bibitem{Crapo} Henry Crapo, {\it Structural Rigidity}, {\it Structural
Topology} {\bf 1} (1979), 26-45.
%
\bibitem{Whiteley} Walter Whiteley, {\it Realizability of Polyhedra},
Structural Topology, {\bf 1} (1979), 46-58.
%
\bibitem{Tesis} H.~Telley, {\it Modelisation et Simulation
Bidimensionelle de la Croissance des Polycrystaux}, Unpublished, Thesis
Report N. 780. Ecole Polytechnique Federale de Lausanne, (1989).
%
\bibitem{vertex} This is called ``phantom vertex'' in \cite{AshBolker2}.
%
\bibitem{Coxeter} {\it Introduction to Geometry}, H.~S.~M.~Coxeter,
Second Edition,  Wiley and Sons 1989.
%
\bibitem{ProjectiveGeometry} A configuration $(p_\gamma, l_\pi)$ is
formed by $p$ points and $l$ lines, such that each line is adjacent to
$\pi$ points, and each point is adjacent to $\gamma$ lines. See for
example, {\it Geometry and the Imagination}, D.~Hilbert and
S.~Cohn-Vossen, Chelsea Publishing Co., New York 1952, chapter III.
%
\bibitem{CFM} C.~Moukarzel,{\it Voronoi Foams}, Physica {\bf A 199}
(1993),19-30.
%
\bibitem{LHM} K.~Lauritsen, C.~Moukarzel and H.~J.~Herrmann, {\it
Statistical Physics and Mechanics of Random Lattices}, {\it
J.~Physique~I} (France) {\bf 3} (1993),1941.
%
\bibitem{CFMtbp} C.~Moukarzel, to be published.
%
\bibitem{equidistance} Within our generalized notion of distance, that
is, including the multiplicative constants. 
%
\bibitem{unstable} Although it is statically unstable, a foam all of
whose films are under compression would also be geometrically possible.
%
\end{references}
\end{document}